%% file: nanograv_9yr_binary.tex
\documentclass[iop]{emulateapj}

\usepackage{apjfonts}
\usepackage{epstopdf}
\usepackage{amsmath}
\usepackage{graphicx}
\usepackage{natbib}
\usepackage{url}
\usepackage[plainpages=false, colorlinks=true, anchorcolor=blue, linkcolor=blue, citecolor=blue, bookmarks=false]{hyperref}

\begin{document}

\title{The NANOGrav Nine-year Data Set: Mass and Geometric Measurements of Binary Millisecond Pulsars}

\author{
Emmanuel~Fonseca\altaffilmark{1},
Timothy~T.~Pennucci\altaffilmark{2},
Justin~A.~Ellis\altaffilmark{3,18},
Ingrid~H.~Stairs\altaffilmark{1,4},
David~J.~Nice\altaffilmark{5},
Scott~M.~Ransom\altaffilmark{6},
Paul~B.~Demorest\altaffilmark{7},
Zaven~Arzoumanian\altaffilmark{8},
Kathryn~Crowter\altaffilmark{1},
Timothy~Dolch\altaffilmark{9,10},
Robert~D.~Ferdman\altaffilmark{11},
Marjorie~E.~Gonzalez\altaffilmark{1,12},
Glenn~Jones\altaffilmark{2},
Megan~L.~Jones\altaffilmark{13},
Michael~T.~Lam\altaffilmark{10},
Lina~Levin\altaffilmark{14},
Maura~A.~McLaughlin\altaffilmark{13},
Kevin~Stovall\altaffilmark{15},
Joseph~K.~Swiggum\altaffilmark{16},
and
Weiwei~Zhu\altaffilmark{17}\\
}

\altaffiltext{1}{Department of Physics and Astronomy, University of British Columbia, 6224 Agricultural Road, Vancouver, BC V6T 1Z1, Canada}
\altaffiltext{2}{Department of Astronomy, Columbia University, 550 W. 120th St. New York, NY 10027, USA}
\altaffiltext{3}{Jet Propulsion Laboratory, California Institute of Technology, 4800 Oak Grove Dr. Pasadena CA, 91109, USA}
\altaffiltext{4}{McGill Space Institute, 3550 University, Montreal, Quebec, H3A 2A7, Canada}
\altaffiltext{5}{Department of Physics, Lafayette College, Easton, PA 18042, USA}
\altaffiltext{6}{National Radio Astronomy Observatory, 520 Edgemont Road, Charlottesville, VA 22903, USA}
\altaffiltext{7}{National Radio Astronomy Observatory, P.~O.~Box 0, Socorro, NM, 87801, USA}
\altaffiltext{8}{Center for Research and Exploration in Space Science and Technology, X-ray Astrophysics Laboratory, NASA Goddard Space Flight Center, Code 662, Greenbelt, MD 20771, USA}
\altaffiltext{9}{Department of Physics, Hillsdale College, 33 E. College Street, Hillsdale, Michigan 49242, USA}
\altaffiltext{10}{Department of Astronomy, Cornell University, Ithaca, NY 14853, USA}
\altaffiltext{11}{Department of Physics, McGill University, 3600 rue Universite, Montreal, QC H3A 2T8, Canada}
\altaffiltext{12}{Department of Nuclear Medicine, Vancouver Coastal Health Authority, Vancouver, BC V5Z 1M9, Canada}
\altaffiltext{13}{Department of Physics, West Virginia University, P.O.  Box 6315, Morgantown, WV 26505, USA}
\altaffiltext{14}{Jodrell Bank Centre for Astrophysics, School of Physics and Astronomy, The University of Manchester, Manchester M13 9PL, UK}
\altaffiltext{15}{Department of Physics and Astronomy, University of New Mexico, Albuquerque, NM, 87131, USA}
\altaffiltext{16}{Department of Physics, University of Wisconsin-Milwaukee, Milwaukee WI 53211, USA}
\altaffiltext{17}{Max-Planck-Institut f{\" u}r Radioastronomie, Auf dem H{\" u}gel 69, D- 53121, Bonn, Germany}
\altaffiltext{18}{Einstein Fellow}

\begin{abstract}
    We analyze 24 binary radio pulsars in the North American Nanohertz Observatory for Gravitational Waves (NANOGrav) nine-year data set. We make fourteen significant measurements of Shapiro delay, including new detections in four pulsar-binary systems (PSRs J0613$-$0200, J2017+0603, J2302+4442, and J2317+1439), and derive estimates of the binary-component masses and orbital inclination for these MSP-binary systems. We find a wide range of binary pulsar masses, with values as low as $m_{\rm p} = 1.18^{+0.10}_{-0.09}\text{ M}_{\odot}$ for PSR J1918$-$0642 and as high as $m_{\rm p} = 1.928^{+0.017}_{-0.017}\text{ M}_{\odot}$ for PSR J1614$-$2230 (both 68.3\% credibility). We make an improved measurement of the Shapiro timing delay in the PSR J1918$-$0642 and J2043+1711 systems, measuring the pulsar mass in the latter system to be $m_{\rm p} = 1.41^{+0.21}_{-0.18}\text{ M}_{\odot}$ (68.3\% credibility) for the first time. We measure secular variations of one or more orbital elements in many systems, and use these measurements to further constrain our estimates of the pulsar and companion masses whenever possible. In particular, we used the observed Shapiro delay and periastron advance due to relativistic gravity in the PSR J1903+0327 system to derive a pulsar mass of $m_{\rm p} = 1.65^{+0.02}_{-0.02}\text{ M}_{\odot}$ (68.3\% credibility).   We discuss the implications that our mass measurements have on the overall neutron-star mass distribution, and on the ``mass/orbital-period" correlation due to extended mass transfer.
\end{abstract}


\section{Introduction}

Radio pulsars serve as probes of gravitation and binary-formation mechanisms when embedded within different types of orbital systems. Many aspects regarding the evolution of their progenitor orbits can be inferred from precise measurements of the five basic Keplerian parameters and the observed spin properties \citep[see][for a review]{lor08}. Pulsars within relativistic binary systems further exhibit a variety of ``post-Keplerian" (PK) effects that can be used to measure additional parameters of each system, such as the component masses and system orientation. PK measurements offer uniquely powerful constraints on the internal structure of ultra-compact objects \citep[e.g.][]{lp04} and the inferred mass distribution of the neutron-star population \citep{tc99,opns12,kkdt13}. Binary radio pulsars therefore provide a desirable environment to test gravitational theory and understand late-stage stellar evolution with high precision. 

Millisecond pulsars (MSPs) -- with exceptionally stable rotation periods $P < 20$ ms and spin-down rates $\dot{P} < 10^{-17}$ -- are understood to be the byproducts of prolonged, stable mass transfer onto a neutron star from an evolving (sub)giant progenitor companion. This long-term recycling process due to Roche-lobe overflow increases the neutron star's spin frequency while circularizing its orbit over the course of accretion. The resultant companion will likely be a low-mass white dwarf (WD), but it is possible for it to be fully evaporated by the high-energy radiation from the spun-up neutron star \citep{rst89a}. Moreover, the dissipative tidal interactions due to stable mass transfer between components will govern the dynamical evolution of the orbit up to the termination of transfer \citep[e.g.][]{phi92, ts99a}; the post-accretion orbital elements will therefore depend on several accretion-related factors, such as the evolving thermal response of the donor star. A notable prediction from binary-evolution theory is a correlation between the resultant mass of the WD companion ($m_{\rm c}$) and post-accretion orbital period ($P_{\rm b}$) for wide binary systems \citep[with $P_{\rm b} > 1$ day; e.g.][]{ts99a}. Evolutionary models can therefore be used in conjunction with pulsar-timing measurements to constrain additional parameters of interest, such as the pulsar mass ($m_{\rm p}$) and the inclination of the orbit relative to the plane of the sky ($i$).

Mass and geometric parameters can be inferred from measurement of PK orbital elements. For example, \cite{sha64} showed that electromagnetic radiation experiences a time delay as it passes near a massive body due to strong-field gravitation. In MSP-binary systems, the pulsed signal will periodically traverse different amounts of spacetime curvature as the pulsar passes in front of or behind its binary companion in its orbit, relative to a distant observer. According to the theory of general relativity (GR), the observed Shapiro timing delay depends on companion mass and the degree of inclination of the binary system, and so a significant measurement of this effect alone yields estimates of both parameters simultaneously. Furthermore, relativistic binary systems exhibit additional PK effects and secular variations of the orbital element due to strong gravitational fields generated by both binary components \citep[e.g.][]{sta03}. Pulsars found in such systems have been used to place formidable constraints on departures from GR in cases where the binary companions are other neutron stars \citep{ksm+06, wnt10, fst14} or WDs \citep{fwe+12, afw+13}. 

The NANOGrav collaboration makes high-precision timing observations of an array of MSPs with the goal of detecting and characterizing gravitational waves at nanohertz frequencies.  Pulsar timing solutions developed for this project contain a wealth of astrophysical information regarding spin, astrometric, orbital, and line-of-sight interstellar medium properties for each MSP.  The NANOGrav nine-year data set \citep{abb+15b} includes measurements and timing solutions of 37 MSPs over time spans as long as nine years, where 25 of these pulsars reside in binary systems. In this paper, we analyze the orbital parameters of 24 binary MSP systems in the NANOGrav timing array; the 25th NANOGrav binary MSP, PSR J1713+0747, was separately studied by \citet{zsd+15} using NANOGrav and historic pulsar data.

In Section \ref{sec:obs}, we provide details regarding the general NANOGrav observing program as well as targeted observations that were obtained specifically for the detection of possible Shapiro timing delays in several NANOGrav MSP-binary systems. In Section \ref{sec:analysis}, we describe the timing models and analytical methods used to derive the orbital elements, as well as theoretical constraints that can be placed on the component masses and system orientation from observed variations in the orbital elements. In Section \ref{sec:results}, we discuss the methods used to characterize the physical parameters of interest, and in particular the component masses and system geometries. In Section \ref{sec:discussion}, we discuss results obtained for individual MSP-binary systems that exhibit Shapiro delays and/or new measurements of secular variations. Finally, in Section \ref{sec:conclusions}, we summarize the main findings of our study and provide a broader context for the implications these measurements have on understanding stellar-binary evolution and the overall mass distribution of binary MSPs.

\section{Observations \& Reduction}
\label{sec:obs}

The full details regarding data collection, calibration, pulse arrival-time determination and noise modeling for the NANOGrav nine-year data set are provided in \citet{abb+15b}. Here we provide a brief summary of this information. The data are publicly available for download online.\footnote{\url{http://data.nanograv.org}}

All 37 NANOGrav MSPs were observed on a monthly basis using either the 300-m William E. Gordon Arecibo Telescope in Puerto Rico or the 100-m Robert C. Byrd Green Bank Telescope (GBT) in West Virginia, USA, as early as 2004 until late 2013. In the cases of PSRs J1713+0747 and B1937+21, both telescopes were used to monitor these MSPs. In addition to the monthly-cadence program, concentrated observing campaigns of 12 MSPs were made at specific orbital phases and were designed to maximize sensitivity to the Shapiro timing delay \citep{pen15}.

For the monthly observations at both telescopes, as well as the targeted Shapiro-delay campaigns at Arecibo, each MSP was observed using two radio receivers at widely separated frequencies in order to accurately measure the pulsar's line-of-sight dispersion properties on monthly timescales, and to account for any evolution in these frequency-dependent properties over time. The dual-receiver observations at Arecibo were performed contiguously during each observing session. The same measurements at the GBT were typically performed within several days of one another due to a need for retraction and extension of the prime-focus boom when switching between receivers. For the targeted Shapiro-delay observations at the GBT, only one receiver was used due to time constraints. The receivers used for the NANOGrav observations reported here were centered near: 327 MHz (at Arecibo only); 430 MHz (at Arecibo only); 820 MHz (at GBT only); 1400 MHz; and 2030 MHz (at Arecibo only).

Two generations of pulsar backend processors were used at each telescope for real-time coherent de-dispersion and folding of the signal using pre-determined ephemerides of each MSP based on early timing solutions. The identical ASP and GASP pulsar machines \citep{dem07,fer08} were used from the start of the NANOGrav observing program in 2004 until their decommissioning in 2011-2012. These backends decomposed the incoming signal into contiguous 4-MHz channels that spanned 20-64 MHz in usable bandwidth, depending on the receiver used and radio-frequency-interference environment. The PUPPI and GUPPI machines \citep{drd+08,fdr10}, currently in use at both telescopes, can process up to 800 MHz in bandwidth using smaller, 1.5625-MHz channels. Both sets of machines generated folded pulse profiles resolved into 2048 bins across the pulsar's spin period.

\citet{abb+15b} used the standard cross-correlation method for the determination of each folded profile's time of arrival (TOA), where a single, de-noised profile template is matched in the Fourier domain with all profiles obtained at some observing frequency and bandwidth \citep{tay92}. Prior to correlation, we averaged data both over time (20-30 min or 2\% of a MSP-binary orbit per TOA, whichever was shorter) and over a small fraction of the available bandwidth.

\section{Binary Timing Models}
\label{sec:analysis}

We used the TEMPO2 pulsar-timing software package\footnote{\url{http://sourceforge.net/projects/tempo2/}} \citep{hem06} for the analysis of topocentric TOAs collected for all NANOGrav binary MSPs, based on the solutions made publicly available by \citet{abb+15b}. Each timing model includes parameters that describe the given pulsar's spin and spin-down rates, astrometry (i.e. ecliptic-coordinate position, proper motion and annual timing parallax), time-varying dispersion measure (DM), binary motion, and evolution in pulse-profile structure as a function of observing frequency. 

For each binary system, five Keplerian parameters were included in the timing model. We also included timing parameters that describe secular variations in the orbital elements, and/or the Shapiro timing delay, if the least-squares fit in TEMPO2 was significantly improved, such that the F-test significance value was at least 0.0027 (i.e. each parameter is at least 3$\sigma$ significant). Finally, we chose to fit for secular variations in the projected semi-major axis ($x$) for PSRs J1600$-$3053 and J1909$-$3744, and a secular variation in $P_{\rm b}$ of PSR J1614$-$2230, despite their lack of 3$\sigma$ significance; the reasons for these additions are discussed in Section \ref{sec:discussion} below. 

We adopted the following definitions of orientation for all MSP-binary systems analyzed in this work: the system inclination angle lies within a range of values $0^{\circ} < i < 180^{\circ}$, with $i = 0^{\circ}$ corresponding to the orbital angular momentum vector pointing in the direction of the Earth; the longitude of the ascending node ($\Omega$) lies within the range $0^{\circ} < \Omega < 360^{\circ}$ and is measured from celestial North through East. The above  definitions are in agreement with the standard astronomical convention and are adopted by TEMPO2 \citep{ehm06}.

\subsection{Keplerian Timing Models}
\label{subsec:models}

The five Keplerian elements for each MSP-binary system were fitted using either the pulsar-timing binary model developed by \citet[][``DD"]{dd85, dd86} or the model of \citet[][``ELL1"]{lcw+01}. Both models use $P_{\rm b}$ and $x$ as timing-model parameters. The DD model is a general description of an orbit with eccentricity $e$ and a well-defined location of periastron, such that the longitude ($\omega$) and time ($T_0$) of periastron passage can be measured with numerical stability. The ELL1 model is applicable for orbits with very small eccentricities, where $\omega$ and $T_0$ are highly covariant. This low-$e$ model instead parametrizes the orbit with $x$,  $P_{\rm b}$, the ``Laplace-Lagrange" parameters ($\eta$, $\kappa$), and time of passage through the ascending node ($T_{\rm asc}$). Both the DD and ELL1 models contain fittable secular-variation parameters that  describe PK and/or geometric phenomena.

The Keplerian elements of the NANOGrav binary MSPs, first published by \citet{abb+15b}, are shown in Table \ref{tab:binarypar}. For NANOGrav MSPs that used the ELL1 binary timing model in the nine-year data release, we derived the DD parameters and their corresponding uncertainties using the relations derived by \citet{lcw+01} in order to show the degree to which eccentricities are well measured.

\subsection{Parametrizations of the Shapiro Delay}
\label{subsec:SDmodel}

The relativistic Shapiro timing delay ($\Delta_{\rm S}$) is incorporated into both the DD and ELL1 binary models, with the low-eccentricity expansion of $\Delta_{\rm S}$ implemented into the latter formalism. In both timing models, $\Delta_{\rm S}$ is a function of the ``range" ($r$) and ``shape" ($s$) parameters, where $s = \sin i$ in most theories of gravitation, including GR. According to GR, $r = {\rm T}_{\odot}m_{\rm c}$, where ${\rm T}_{\odot} = G{\rm M}_{\odot}/c^3 = 4.925490947$\ $\mu s$ is a conversion factor to units of solar mass. In what follows below, we assume GR is correct in order to compute estimates of $m_c$ from measured Shapiro-delay signals. The original analysis of the NANOGrav nine-year data set used only the traditional ($m_{\rm c}$, $\sin i$) parameterization of the Shapiro delay, incorporating one or two Shapiro parameters if they met the F-test criteria described earlier. 

For this detailed study of MSP-binary systems, we also created timing solutions that used the ``orthometric" parametrization of the Shapiro timing delay \citep{fw10}. The orthometric framework reparameterizes $\Delta_{\rm S}$ as a Fourier expansion across each system's orbital period and uses two different PK parameters that are derived from the harmonics of the Shapiro-delay signal to describe the relativistic effect. In the orthometric framework, the PK parameters are either the third and fourth harmonic amplitudes of $\Delta_{\rm S}$ ($h_3$ and $h_4$, respectively), or $h_3$ and the orthometric ratio $\varsigma = h_4/h_3$. The choice of ($h_3$, $h_4$) as PK parameters is most appropriate for low-$e$  systems with $i < 60^{\circ}$, while the ($h_3$, $\varsigma$) combination is best used for eccentric systems and low-$e$ systems with $i > 60^{\circ}$. 

While no new physical information is made available by its PK parameters, the orthometric parametrization reduces correlation between the Shapiro-delay parameters. The orthometric model therefore provides a more numerically stable solution to the timing of binary pulsars with significant Shapiro-delay signals, particularly in low-$e$, low$-i$ systems where $\Delta_{\rm S}$ is difficult to measure. The available orthometric PK parameters are related to the traditional PK parameters as nonlinear functions:

\begin{align}
    \label{eq:stig}
    \varsigma &= \sqrt{\frac{1-\cos i}{1+\cos i}}\\
    \label{eq:h3}
    h_3 &= r\varsigma^3 \\
    \label{eq:h4}
    h_4 &= h_3\varsigma.
\end{align}

As shown by \citet{fw10}, the statistical significance of $h_3$ reflects the degree to which $\Delta_{\rm S}$ is measurable and can therefore be used as a straightforward indicator for the detection of the Shapiro timing delay in a pulsar-binary system. In this study, we considered the Shapiro delay to be measurable if the estimate of $h_3$ was statistically significant to at least $3\sigma$. For all systems with significant $\Delta_{\rm S}$, as well as systems with statistically significant eccentricities that did not pass the $h_3$ significance test, we used the $(h_3, \varsigma)$ parameters to describe the Shapiro timing delay. For low-$e$ systems with no significant $\Delta_{\rm S}$, we instead parameterized $\Delta_{\rm S}$ using the ($h_3$, $h_4$) combination.

Given the relations between the ($m_{\rm c}$, $\sin i$) and ($h_3$,$\varsigma$) parameters in Equations \ref{eq:stig} and \ref{eq:h3}, physical arguments require that $h_3 > 0$ and $0 < \varsigma < 1$. Equation \ref{eq:h4} subsequently requires that $h_4$ be positive, as well.  However, TEMPO2 does not impose any theoretically-motivated constraints on the Shapiro-delay parameters (traditional or orthometric) during a timing-model fit; it is therefore mathematically allowed for the Shapiro-delay terms to possess values that exceed their physical limits. Such limit discrepancies are not expected to be an issue for significant $\Delta_{\rm S}$ signals, but may occur for non-detections of the Shapiro delay due to large statistical correlation between parameters when the $\Delta_{\rm S}$ signal is weak.

\subsection{Secular and Periodic Variations of Orbital Elements}
\label{subsec:secvar}

Pulsar-binary systems in tight orbits with WDs or other neutron stars typically exhibit PK effects that are observed as secular changes in the orbital elements. In order to analyze these effects, we assumed GR to be valid; we refer to the secular PK variations in this study as $(\dot{P_{\rm b}})_{\rm GR}$, $(\dot{\omega})_{\rm GR}$, $(\dot{x})_{\rm GR}$ and $(\dot{e})_{\rm GR}$, where the dots indicate derivatives in time. According to GR, each of the PK quantities (including the Shapiro $r$ and $s$ parameters) are functions of at least one of the two component masses \citep{dt92}. We neglected the $(\dot{x})_{\rm GR}$ and $(\dot{e})_{\rm GR}$ terms since these two quantities are negligible for MSP-binary systems on the timescale of the NANOGrav nine-year data set. We considered the GR terms for $\dot{P_{\rm b}}$, $\dot{x}$, and/or $\dot{\omega}$ in the interpretation of observed variations in PSRs J1600$-$3053, J1614$-$2230, J1903+0327, and J1909$-$3744.

Besides the intrinsic changes within orbits from PK effects, apparent secular variations in the orbital elements will also be induced from significant relative motion between the MSP-binary and Solar-system-barycentre (SSB) reference frames \citep{kop96}. The secular variations in $x$ and $\omega$ from proper motion ($\mu$) -- to which we refer in this study as $(\dot{x})_{\mu}$ and $(\dot{\omega})_{\mu}$ -- arise from a long-term change in certain elements of orientation as the binary system moves across the sky. The kinematic terms for $\dot{x}$ and $\dot{\omega}$ are described as trigonometric functions of $i$ and $\Omega$. We considered the kinematic terms for $\dot{x}$ and/or $\dot{\omega}$ in the interpretation of observed variations in PSRs J1455$-$3330, J1600$-$3053, J1640+2224, J1643$-$1224, J1741+1351, J1853+1303, B1855+09, J1909$-$3744, J1910+1256, B1953+29, J2145$-$0750, and J2317+1439.

A separate kinematic bias that produces observed secular variations in orbital elements can arise from several forms of relative acceleration between the MSP-binary and SSB systems \citep[e.g.][]{nt95}, the most prominent of which are: differential rotation in the Galactic disk; acceleration in the Galactic gravitational potential vertical to the disk \citep[e.g.][]{kg89}; and apparent acceleration due to significant proper motion \citep{shk70}. The kinematic bias from relative acceleration produces a rate of change in the Doppler shift ($D$) in $P_{\rm b}$ and ultimately produces an apparent variation in the orbital period. We refer to the component of the secular variation due to the acceleration bias as $(\dot{P}_{\rm b})_{D}$, and considered this effect in the interpretation of measured variations in PSRs J1614$-$2230 and J1909$-$3744.

For sufficiently nearby pulsar-binary systems, the observed system orientation will change periodically as the Earth and the MSP orbit their respective barycenters and at their respective orbital periods. The ``mixed" periodic variations in $x$ and $\omega$, collectively referred to as the ``annual orbital parallax" \citep{kop95}, depend on $i$, $\Omega$, and the observed parallax ($\varpi$) of the pulsar-binary system. Annual orbital parallax has been measured for PSRs J0437-4715 \citep[e.g.][]{vbv+08} and J1713+0747 \citep[e.g.][]{zsd+15} and has been used in conjunction with the reported Shapiro timing delays  and annual astrometric parallaxes to uniquely solve for the three-dimensional geometry of these two binary systems. We considered this effect when analyzing the secular variations observed in PSRs J1640+2224 and J1741+1351.

\section{Analyses of Mass \& Geometric Parameters}
\label{sec:results}

We measured the Shapiro timing delay in fourteen binary-MSP systems, as well as many secular variations, based on the F-test significance criterion used by \citet{abb+15b}. The same fourteen systems with significant $\Delta_{\rm S}$ also passed the 3$\sigma$-significance test of $h_3$, as described in Section \ref{subsec:SDmodel}. The secular/PK measurements are shown in Table \ref{tab:postKeppar}, and the derived estimates of $m_{\rm p}$, $m_{\rm c}$, and $i$ are shown in Table \ref{tab:SDresults}. 

\subsection{Statistical Analyses of Shapiro-Delay Signals}
\label{subsec:grids}

We used the procedure outlined by \citet{sna+02} in order to perform a statistically rigorous analysis of the fourteen MSPs in the nine-year data set with significant Shapiro-delay measurements and obtain robust estimates of $m_{\rm p}$, $m_{\rm c}$, and $i$. For each of the fourteen MSPs, we first created a uniform, two-dimensional $n\times n$ grid of $\chi^2$ values for different combinations of $m_{\rm c} = r/{\rm T}_{\odot}$ and $\cos i$, where $n = 200$ or greater in order to minimize artifacts from interpolation. With the exception of the noise parameters, all other timing-model parameters were allowed to vary freely when estimating the $\chi^2$ at each grid coordinate; the noise terms were held fixed at their maximum-likelihood values as determined by \citet{abb+15b}. We used $\cos i$ instead of $\sin i$ as a grid coordinate since a collection of randomly-oriented binary systems possesses a uniform distribution in $\cos i$.

Each $\chi^2$ map was then converted to a two-dimensional probability distribution function (PDF) by using a likelihood density of the following form,

\begin{equation}
    \label{eq:likelihood}
    p(\text{data}|m_{\rm c},\cos i) \propto e^{-(\chi^2-\chi^2_0)/2}
\end{equation}

\noindent where $\chi^2_0$ is the minimum value of the $\chi^2$ distribution defined on the two-dimensional grid. Bayes' theorem subsequently yields the two-dimensional posterior PDF, $p(m_{\rm c},\cos i|\text{data})$, when using the joint-uniform prior distribution of the two Shapiro-delay parameters. We then marginalized (i.e. integrated) the two-dimensional PDF over $\cos i$ to obtain the one-dimensional PDF in $m_{\rm c}$, and marginalized over $m_{\rm c}$ to obtain the one-dimensional PDF in $\cos i$. In order to obtain a PDF in $m_{\rm p}$, we transformed the two-dimensional ($m_{\rm c}$, $\cos i$) probability grid to one in the ($m_{\rm p}$, $\cos i$) space by applying the transformation rule for PDFs of random variables, 

\begin{equation}
    p(m_{\rm p}, \cos i|{\rm data}) = p(m_{\rm c}, \cos i|{\rm data}) \frac{\partial m_{\rm c}}{\partial m_{\rm p}},
\end{equation}

\noindent where the partial derivative is evaluated by using the mass function (for a fixed value of $\cos i$). 

For the two-dimensional grids, we computed $\chi^2$ values over $0<\cos i<1$ and, unless otherwise noted, $0<m_{\rm c}<1.4$ M$_\odot$. The latter upper limit approximately corresponds to the Chandrasekhar limit for a non-rotating white dwarf. (Three exceptions to this cut-off limit are PSRs J1903+0327, J1949+3106, and J2302+4442, which are discussed individually in Section \ref{sec:discussion} below.)

We applied the same set of $\chi^2$-grid and marginalization procedures described above for the fourteen timing models with significant $\Delta_{\rm S}$ that used the ($h_3$, $\varsigma$) orthometric parametrization. However, we first created a $\chi^2$ grid in uniform steps of the ($h_3$, $\varsigma$) parameters, and afterwards converted the resultant likelihood density to the ($m_c$, $\cos i$) probability map by using Equations \ref{eq:stig} and \ref{eq:h3} when applying the PDF-transformation rule. 

The choice in parametrization of $\Delta_{\rm S}$ amounts to a difference in prior probabilities on the physical parameters ($m_{\rm c}$, $\sin i$) when performing the MCMC or $\chi^2$-grid analysis described above, due to the nonlinear relation between the physical and orthometric parameters (Equations \ref{eq:stig}-\ref{eq:h4}). Our first choice of prior, in ($m_{\rm c}$, $\cos i$), is motivated by the expected distribution of randomly oriented binary systems -- uniform in $\cos i$ -- though the choice of uniform $m_{\rm c}$ is arbitrary. On the other hand, \citet{fw10} argue that a statistical analysis of the orthometric parameters is preferable since $h_3$ and $\varsigma$ are related to the Fourier harmonics of $\Delta_{\rm S}$ and make no immediate assumption on the probability distributions of physical parameters. Simulations by Freire \& Wex show that the one-dimensional posterior PDFs of the physical parameters will be affected in cases of low inclination, where $\Delta_{\rm S}$ is typically weaker and the posterior density is heavily influenced by the choice of prior information. For cases in which there is a highly-significant measurement of $\Delta_{\rm S}$, such that the posterior density spans a small range of parameter space, the two choices of priors give essentially the same results. We present the results obtained from both sets of priors to demonstrate the effects such choices have on our mass measurements.

\subsubsection{MCMC Analysis of Shapiro-delay Parameters}
\label{subsubsec:gridMCMCcomp}

As a check on the $\chi^2$-grid procedure described above, we evaluated the parameters of each binary system  using a Bayesian Markov Chain Monte Carlo \citep[MCMC; e.g.][]{gre05bayes} analysis of all timing-model parameters. In the MCMC analysis, where we used the PAL2 Bayesian inference suite\footnote{\url{https://github.com/jellis18/PAL2}}, the joint likelihood density includes all spin, astrometric, binary and noise terms as parameters to be sampled. The Bayesian analysis uses the traditional ($m_{\rm c}$, $\cos i$) parameterization for the Shapiro delay, along with uniform priors on these and all other timing model parameters. We analytically marginalized the joint posterior over the DM, profile-evolution, and backend-offset parameters in order to reduce computational needs.

In principle, the MCMC analysis therefore provides a more robust exploration of the parameter space and timing-model behavior than the $\chi^2$-grid analysis, since the MCMC method samples the noise parameters, while the $\chi^2$-grid holds the noise parameters fixed. Moreover, for the MCMC analysis, the computation of $m_{\rm p}$ accounts for the small uncertainty in the mass function, as it uses the posterior distributions for the Shapiro-delay and Keplerian parameters.

\begin{figure}
    \centering
    \includegraphics[scale=0.5]{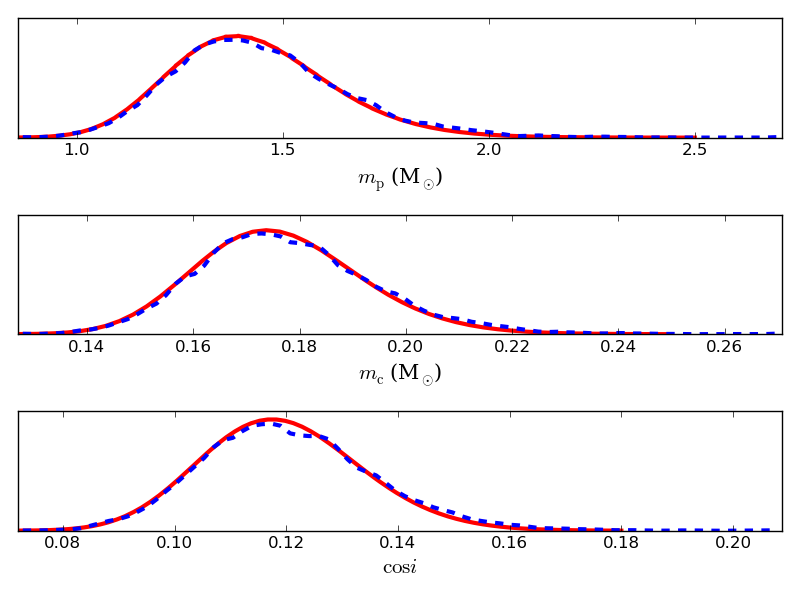}
    \caption{Normalized posterior PDFs of $m_{\rm p}$, $m_{\rm c}$ and $\cos i$ for PSR J2043+1711. The red-solid curves were obtained from a $\chi^2$-grid analysis, and the blue-dashed curves were generated from an MCMC analysis of all timing-model parameters (including terms that characterize red- and white-noise processes) when drawing $10^6$ samples and using a thinning factor of 10 to reduce autocorrelation. The $\chi^2$-grid and MCMC methods yield nearly identical estimates of the posterior PDFs.}
    \label{fig:gridMCMCcomp}
\end{figure}

Figure \ref{fig:gridMCMCcomp} shows the normalized posterior PDFs of the Shapiro-delay parameters for PSR J2043+1711 (see Section \ref{subsec:J2043}) estimated from both the $\chi^2$-grid and MCMC analyses. It is clear that the $\chi^2$-grid and MCMC analyses yield nearly identical estimates of the posterior distributions of the component masses and $\cos i$. This consistency between methods is seen for all 14 MSPs with significant $\Delta_{\rm S}$. Thus the $\chi^2$-grid method is a reliable method for estimating posterior PDFs when using an adequate (fixed) noise model. All estimates reported below were obtained from the $\chi^2$-grid method and verified using PAL2.

\subsubsection{Constraints from $(\dot{\omega})_{\rm GR}$ on Shapiro-delay Parameters}
\label{subsubsec:OMDOTconstraints}

Both PSRs J1600$-$3053 and J1903+0327 exhibit statistically significant measurements of $\dot{\omega}$ and $\Delta_{\rm S}$. As discussed in Sections \ref{subsec:J1600} and \ref{subsec:J1903} below, the $\dot{\omega}$ measurements in these two systems are likely due to GR. We therefore generated additional $\chi^2$ grids of the two Shapiro-delay parameters for PSRs J1600$-$3053 and J1903+0327 that used the statistical significance of $\dot{\omega}$ to improve our estimates of the Shapiro-delay parameters in the following manner:

\begin{itemize}
    \item for each ($m_{\rm c}$, $\cos i$) coordinate on the $\chi^2$ grid, we computed a value of $m_{\rm p}$ using the mass function for the given system; for the orthometric grids, we first used Equations \ref{eq:stig} and \ref{eq:h3} to compute $m_{\rm c}$ and $\cos i$ at each ($h_3$, $\varsigma$) grid coordinate, and then used the mass function to compute $m_{\rm p}$;
    \item we then used the values of $m_{\rm p}$ and $m_{\rm c}$, along with the Keplerian elements of the given system, to compute $(\dot{\omega})_{\rm GR}$ at the ($m_{\rm c}$, $\cos i$) or ($h_3$, $\varsigma$) grid points;
    \item we then held the $\dot{\omega}$ parameter fixed in the timing solution at the value given by $(\dot{\omega})_{\rm GR}$, along with the Shapiro-delay parameters, and used TEMPO2 to obtain a constrained $\chi^2$ value.

\end{itemize}

\noindent We then used Equation \ref{eq:likelihood} and the marginalization procedures discussed above to obtain constrained PDFs of $m_{\rm p}$, $m_{\rm c}$ and $i$ from both parametrizations of $\Delta_{\rm S}$.

\subsubsection{Constraints from geometric variations on Shapiro-delay Parameters}
\label{subsubsec:XDOTconstraints}

PSRs J1640+2224 and J1741+1351 have significant measurements of $\Delta_{\rm S}$ and secular variations in $x$ that are likely due to proper motion. However, PSR J1640+2224 also exhibits a significant $\dot{\omega}$ that is currently not well understood in terms of the various mechanisms outlined in Section \ref{subsec:secvar} above (see Section \ref{subsec:J1640} for a discussion). We therefore only analyze the observed geometric variation in $x$ for PSR J1741+1351. 

In the case of PSR J1741+1351, we generated $\chi^2$ grids that explicitly modeled the observed $\dot{x}$ in terms of system geometry at each grid point. We used the T2 binary timing model in TEMPO2, a general binary framework that uses the DD or ELL1 models when appropriate but also allows for $i$ and $\Omega$ to be used as fit parameters; the T2 timing model computes both the secular and periodic variations in $x$ and $\omega$ given the two geometric parameters.  

The explicit modeling of orbital variations from geometric biases introduces $\Omega$ as an {\it a priori} unknown parameter; we therefore generated three-dimensional $\chi^2$ grids in the uniform ($m_{\rm c}$, $\cos i$, $\Omega$) and ($h_3$, $\varsigma$, $\Omega$) phase spaces for PSRs J1640+2224 and J1741+1351, using Equation \ref{eq:likelihood} as the Bayesian likelihood at each grid point in the three-dimensional phase space. We then appropriately translated and marginalized the three-dimensional probability maps in order to obtain one-dimensional posterior PDFs of $m_{\rm p}$, $m_{\rm c}$, $i$ and $\Omega$. 

If only one geometric variation is measured, the ($m_{\rm c}$, $\cos i$, $\Omega$) and ($h_3$, $\varsigma$, $\Omega$) grid analyses will introduce a sign ambiguity in $\Omega$ due to the sign ambiguity in $i$ as determined from the Shapiro timing delay. The ambiguity in $\Omega$ results in a four-fold ambiguity in the system orientation ($i$, $\Omega$) of the orbit. However, if two or more secular and/or periodic variations are measured, then the four-fold degeneracy can be broken to determine a unique orientation of the MSP-binary orbit. We consider the relevance of annual orbital parallax for PSRs J1640$+$2224 and J1741$+$1351 below. 

\subsection{Limits on inclination from $\dot{x}$ and the absence of Shapiro Delay}

A constraint on the system inclination angle can still be placed using the $\dot{x}$ measurements listed in Table \ref{tab:postKeppar} \citep[e.g.][]{nss01} for cases where the Shapiro timing delay is not detected. This is possible since the trigonometric term for $\Omega$  cannot exceed unity \citep[Equation 11 in][]{kop96}, which corresponds to an alignment between the proper-motion vector and the projection of the orbital angular moment vector on the plane of the sky. The ``magnitude" of the effect can therefore be written as $|\dot{x}|_{\mu, \rm max} = \mu x |\cot i|$, and an upper limit on the system inclination can be calculated as

\begin{equation}
    i < \arctan\bigg[{\frac{x\mu}{|\dot{x}|_{\rm obs}}}\bigg].
    \label{eq:i95limit}
\end{equation}

\noindent We computed a 95.4\%-credible  upper limit on the system inclination using Equation \ref{eq:i95limit} and the 2$\sigma$ lower limit of the $\dot{x}$ measurements reported in Table \ref{tab:upperlim} for systems with no detected Shapiro delay.

Another constraint on the system inclination can be placed by using a non-detection of the Shapiro timing delay. The Shapiro-delay $\chi^2$ grids of pulsar-binary systems with no measurable $\Delta_{\rm S}$ contain zero probability in regions of the ($m_{\rm c}$, $\cos i$) space that correspond to large companion masses and high inclinations. These regions can be excluded based on statistically poor timing-model fits to the NANOGrav nine-year data sets. 

\begin{figure}
    \centering
    \includegraphics[scale=0.43]{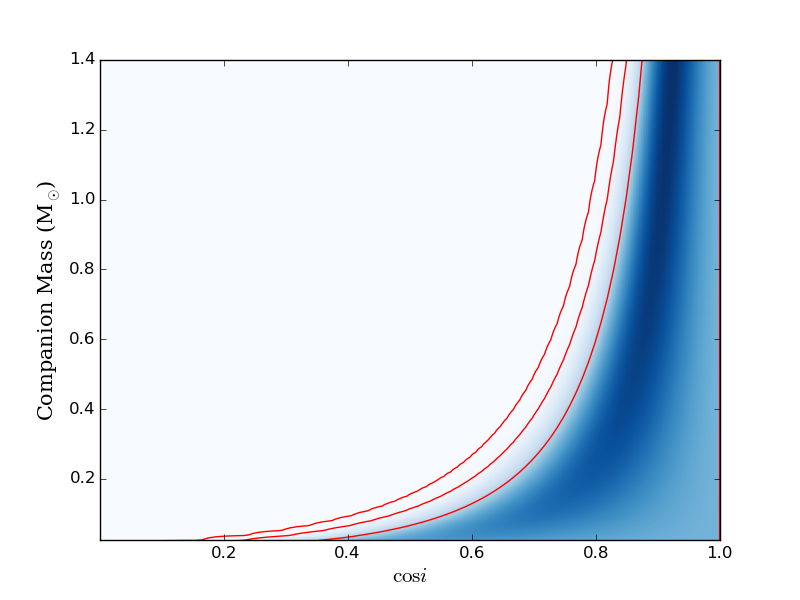}
    \includegraphics[scale=0.43]{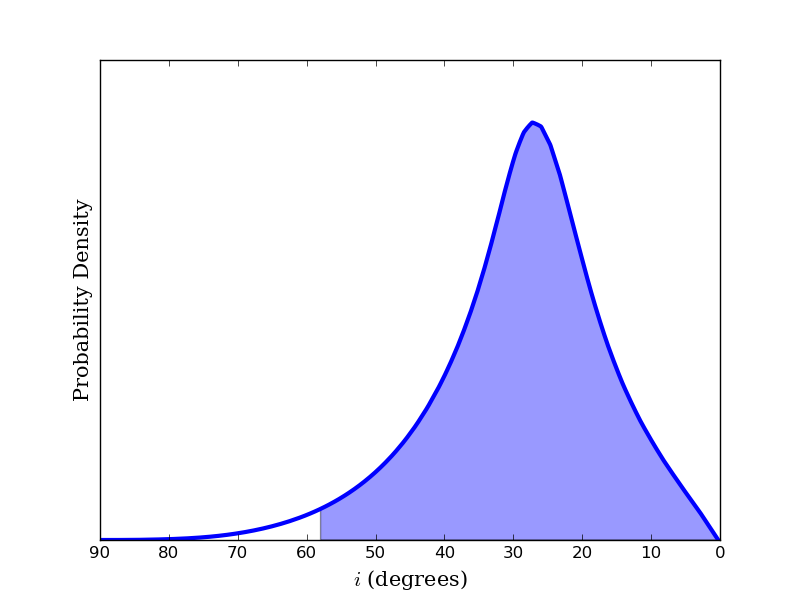}
    \caption{{\it Left.} A ($m_{\rm c}$, $\cos i$) probability map for PSR J0023+0923. The inner, middle and outer red contours encapsulate 68.3\%, 95.4\% and 99.7\% of the total probability. {\it Right.} Posterior PDF of the derived inclination angle for PSR J0023+0923, obtained from  the ($m_{\rm c}$, $\cos i$) grid shown on the left. The shaded blue region under the PDF  contains 95\% of the total probability relative to no inclination.}
    \label{fig:upperlimJ0023}
\end{figure}

However, a complication in the limit on $i$ by using $\chi^2$ grids arises from the cut-off value in $m_{\rm c}$ when generating the $\chi^2$ grids as discussed in Section \ref{subsec:grids}: the cut-off value prevents an arbitrarily large weight in probability density being assigned to values of low $\cos i$ and high $m_{\rm c}$, but also disregards regions of the ($m_{\rm c}$, $\cos i$) phase space with non-zero probability density. We believe that the  cut-off value in $m_{\rm c}$ is nonetheless justified since the only MSP with a suspected main-sequence-star companion is PSR J1903+0327. Moreover, the inclusion of more probability density would shift the upper limit on $i$ to slightly lower values, so the upper limits we report in this study are considered to be conservative. Figure \ref{fig:upperlimJ0023} shows the Bayesian-gridding and upper-limit results for PSR J0023+0923, and the upper limits for NANOGrav binary MSPs with $\dot{x}$ measurements and/or no detections of the Shapiro timing delay are provided in Table \ref{tab:upperlim}. 

\section{Results \& Discussion}
\label{sec:discussion}

The traditional and orthometric parameterizations of the Shapiro timing delay yield consistent measurements of the component masses, $i$, and $\Omega$ (when the latter angle is measurable) in the fourteen NANOGrav MSP-binary systems with significant $\Delta_{\rm S}$ that we analyze here. We report estimates that were made using both Shapiro-delay models for each of these 14 MSPs in Table \ref{tab:SDresults}. Any differences in the estimates and credible intervals derived from the traditional ($m_c$, $\sin i$) or orthometric ($h_3$, $\varsigma$) probability grids reflect different priors on those PK parameters; the most highly-inclined systems produced essentially identical estimates. These features are consistent with the expectations discussed in Section \ref{subsec:grids}.

Unless otherwise specified, all numerical values with uncertainties presented below reflect 68.3\% equal-tailed credible intervals; that is, we compute the credible interval by numerically integrating each (normalized) posterior PDF to values of the parameter that contain 15.9\% (lower bound), 50\% (median), and 84.1\% (upper bound) of all probability.

\subsection{PSR J0613$-$0200}

PSR J0613$-$0200 is a 3.1-ms pulsar in a 1.2-day orbit that was discovered in a survey of the Galactic disk using the Parkes radio telescope \citep{lnl+95}. A previous long-term timing study of this MSP by \citet{hbo06} used the lack of a Shapiro-delay detection to place constraints on the companion mass and system inclination, such that $0.13 < m_{\rm c}/{\rm M}_{\odot} < 0.15$ and $59^{\circ} < i < 68^{\circ}$ if $m_{\rm p} = 1.3\textrm{ M}_{\odot}$. Two recent, independent TOA analyses of PSR J0613-0200 were performed by \citet{rhc+16} and \citet{dcl+16}. Reardon et al. used an 11-yr data set collected for the Parkes Pulsar Timing Array (PPTA) and did not report any secular variations or PK effects. Desvignes et al. used a 16-yr data set collected for the European Pulsar Timing Array (EPTA) to be measure a significant $\dot{P}_{\rm b} = 4.8(1.1)\times10^{-14}$. Neither study reports a detection of the Shapiro timing delay. A recent optical-spectroscopy study did not detect the companion to PSR J0613$-$0200, and placed a 5$\sigma$-detection lower limit on the photometric R-band magnitude to be $R > 23.8$ \citep{bac+15}.

\begin{figure}
    \centering
    \includegraphics[scale=0.45]{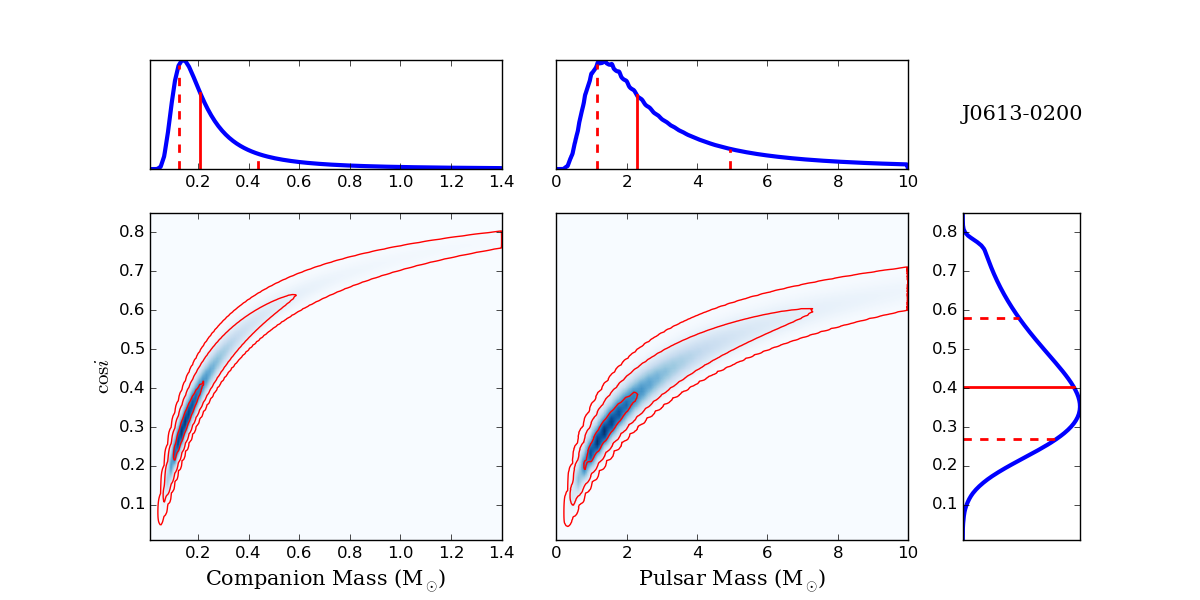}
    \includegraphics[scale=0.45]{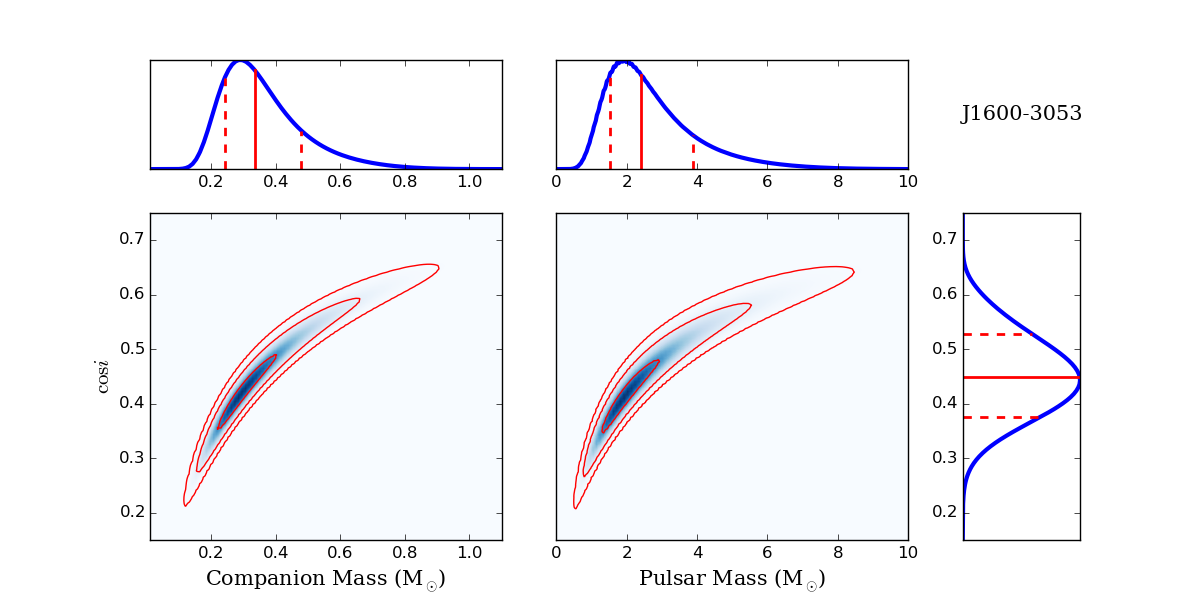}
    \includegraphics[scale=0.45]{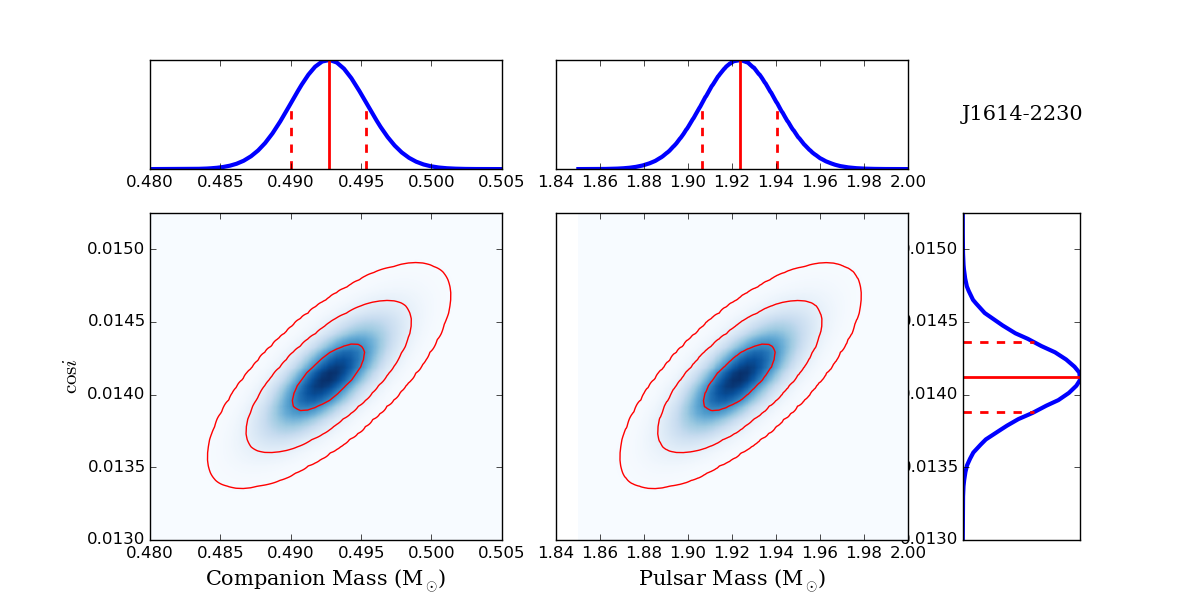}
    \caption{Probability maps and posterior PDFs of the traditional Shapiro-delay parameters measured for PSRs J0613$-$0200, J1600$-$3053, and J1614$-$2230. The maps and PDFs for J1600$-$3053 were constrained assuming that the observed $\dot{\omega}$ is due to GR (see Section \ref{subsec:J1600}). The inner, middle and outer red contours encapsulate 68.3\%, 95.4\% and 99.7\% of the total probability defined on each two-dimensional map, respectively. In all slimmer panels, the blue solid lines represent posterior PDFs obtained from marginalizing the appropriate two-dimensional map, the vertical red-dashed lines are bounds of the 68.3\% credible interval, and the red-solid line is the median value.}
    \label{fig:SDfig1}
\end{figure}

For the first time, we report the detection of the Shapiro timing delay in the PSR J0613$-$0200 system using the NANOGrav nine-year data set. It is likely that the Shapiro-delay signal in PSR J0613$-$0200 went undetected by \citet{rhc+16} and \citet{dcl+16} because of the better sensitivity achieved with the GBT and GUPPI backend, as reflected by the factor of 2-3 improvement in TOA root-mean-square (RMS) residuals between the NANOGrav and PPTA/EPTA data sets. The $\chi^2$ grids and marginalized PDFs for PSR J0613$-$0200 are shown in Figure \ref{fig:SDfig1}. Our current estimates of $m_{\rm c} = 0.18^{+0.15}_{-0.07}\text{ M}_{\odot}$ and $i = 68^{+7}_{-10}$ degrees are consistent with the predictions made by \citet{hbo06}, though our derived estimate of $m_{\rm p} = 2.3^{+2.7}_{-1.1}\text{ M}_{\odot}$ is not yet precise enough to yield a meaningful constraint on the pulsar mass.

\subsection{PSR J1455$-$3330}

PSR J1455$-$3330 is a 7.9-ms pulsar in a 76-day orbit and was discovered in a survey of the Galactic disk using the Parkes radio telescope \citep{lnl+95}. The long spin period of this MSP, along with its large orbit and anomalously large characteristic age, indicates potential disk instability during the transfer phase that ultimately dontated little mass to the neutron star \citep{lvw98}. A recent radio-timing analysis by \citet{dcl+16} reported a significant $\dot{x} = -1.7(4)\times10^{-14}$.

We measured a significant $\dot{x} = -2.1(5)\times10^{-14}$ in the PSR J1455$-$3330 system using the NANOGrav nine-year data set. Our estimate of $\dot{x}$ is consistent with the one made by \citet{dcl+16} using an independent data set. We did not detect a Shapiro timing delay, as indicated by the insignificance of $h_3$ and unconstrained estimate of $\varsigma$ listed in Table \ref{tab:postKeppar}. 

\subsection{PSR J1600$-$3053}
\label{subsec:J1600}

PSR J1600$-$3053 is a 3.6-ms pulsar in a 14.3-day orbit that was discovered in a survey of high Galactic latitudes using the Parkes radio telescope \citep{jbo+07}. A recent analysis of the PSR J1600$-$3053 system by \citet{rhc+16} used PPTA data to make significant measurements of $\dot{x}$ and the Shapiro timing delay: $m_{\rm p} = 2.4(1.7)\text{ M}_{\odot}$, $m_{\rm c} = 0.34(15)\text{ M}_{\odot}$, $\sin i = 0.87(6)$, and $\dot{x} = -4.2(7) \times 10^{-15}$. Another recent and independent study by \citet{dcl+16} used EPTA data to measure the orthometric parameters $h_3 = 0.33(2)\text{ }\mu$s and $\varsigma = 0.68(5)$, consistent with the component masses and inclination measured by Reardon et al., as well as $\dot{x} = -2.8(5)\times10^{-15}$.

We measured a significant $\dot{\omega}$ for the first time, as well as a Shapiro timing delay in the PSR J1600-3053 system. We do not yet measure a $3\sigma$ significant $\dot{x}$, likely because the NANOGrav data span for PSR J1600$-$3053 is $\sim$6 yr, several years shorter than the EPTA and PPTA data sets. Nevertheless, we do make a tentative, $\sim$2$\sigma$ detection of $\dot{x} = -1.7(9)\times10^{-15}$ and have elected to include it as a free parameter in our timing solution. Our estimates of $\dot{x}$ and the orthometric parameters, $h_3 = 0.39(3)$ and $\varsigma = 0.62(6)$, are consistent with those made by \citet{dcl+16}.

Our measurement of $\dot{\omega} = 7(2)\times10^{-3}\text{ deg yr}^{-1}$ in the PSR J1600$-$3053 system could, in principle, be due to a combination of physical effects discussed in Section \ref{subsec:secvar}. The maximum amplitude of ($\dot{\omega})_{\mu}$ for PSR J1600$-$3053 is $(\dot{\omega})_{\mu, \rm max} = \mu |\cot i| \sim 10^{-6}\textrm{ deg yr}^{-1}$, which is two orders of magnitude smaller than the uncertainty level for the observed $\dot{\omega}$ in this MSP-binary listed in Table \ref{tab:postKeppar}. Therefore, the observed $\dot{\omega}$ in the PSR J1600$-$3053 system cannot be due to secular variations from proper motion at the current level of precision. 

The predicted GR component of $\dot{\omega}$ of PSR J1600$-$3053 is on the order of $10^{-3}\text{ deg yr}^{-1}$ given the Keplerian parameters of the system shown in Table \ref{tab:binarypar}, the same order of magnitude as our measured value. We therefore used the method described in Section \ref{subsubsec:OMDOTconstraints} to include both $\dot{\omega}$ and the Shapiro-delay parameters when generating the two-dimensional $\chi^2$ grid. The $\chi^2$ grids and marginalized PDFs for PSR J1600$-$3053 are shown in Figure \ref{fig:SDfig1}; the constrained estimates of the component masses and inclination are: $m_{\rm p} = 2.4^{+1.5}_{-0.9}\text{ M}_{\odot}$; $m_{\rm c} = 0.33^{+0.14}_{-0.10}\text{ M}_{\odot}$; and $i$ = 63(5) degrees. Our constrained estimates of the Shapiro delay parameters are consistent with the estimates made by \citet{rhc+16} and \citet{dcl+16}.

\subsection{PSR J1614$-$2230}

PSR J1614$-$2230 is a 3.2-ms pulsar in a 8.7-day orbit with a massive WD companion; this MSP  was discovered in a mid-latitude radio search of unidentified EGRET gamma-ray sources using the Parkes radio telescope \citep{hrr+05, crh+06a}. The PSR J1614$-$2230 system contains one of the most massive neutron stars known, $m_{\rm p} = 1.97(4)\text{ M}_{\odot}$, as determined by a strategic set of observations that were made and used by \citet{dpr+10} to measure the Shapiro timing delay in this highly-inclined binary system. Demorest et al. were able to rule out nearly all models for plausible neutron-star equations of state that invoke significant amounts of exotic matter. Moreover, the PSR J1614$-$2230 system provided early evidence for relatively high ``birth masses" of neutron stars after their formation, and before the onset of mass transfer \citep{tlk11}. 

We made an improved measurement of the Shapiro timing delay in PSR J1614$-$2230 when using the NANOGrav nine-year data set, which includes a subset of the GUPPI data used by \citet{dpr+10}. The $\chi^2$ grids and marginalized PDFs for PSR J0613$-$0200 are shown in Figure \ref{fig:SDfig1}.  The uncertainties in both $m_{\rm c} = 0.493(3)\text{ M}_{\odot}$ and $i = 89.189(14)$ degrees have decreased such that the uncertainty in $m_{\rm p} = 1.928(17)\text{ M}_{\odot}$ is a factor of $\sim$3 less than that made by \citet{dpr+10}. 

Although there was not a formally significant measurement of orbital decay, we nevertheless explored fitting for it.  We measured $(\dot{P}_{\rm b})_{\rm obs}=1.3(7)\times 10^{-12}$.  This is much larger than the component expected from general-relativistic orbital decay, $(\dot{P}_{\rm b})_{\rm GR} =-0.00042 \times 10^{-12}$. Instead, it is attributable to the change in the Doppler shift due to the pulsar motion, as discussed in Section \ref{subsec:secvar}, which predicts $(\dot{P}_{\rm b})_{\rm D}=1.36\times 10^{-12}$ based on the pulsar distance and proper motion.  \citet{mnf+16} used the agreement between $(\dot{P}_{\rm b})_{\rm D}$  and the observed value as a confirmation of the parallax distance to the pulsar.  The precision of $(\dot{P}_{\rm b})_{\rm obs}$ can be improved by extending the observing span backwards using pre-GUPPI archival data published by \citet{dpr+10} and forwards (through future observations); this will eventually provide the most precise means for measuring the distance to this pulsar.

\subsection{PSR J1640+2224}
\label{subsec:J1640}

PSR J1640+2224 is a 3.1-ms pulsar in a 175-day orbit that was discovered in a Arecibo survey of high Galactic latitudes \citep{fcwa95, fwc95}. The companion star in this system was observed using the Palomar 5.1-m optical telescope to have an effective temperature that is consistent with an old He WD \citep{lcf+95}. The first dedicated radio-timing study of the PSR J1640+2224 system reported a tentative detection of the Shapiro timing delay, with $m_{\rm c} = 0.15^{+0.08}_{-0.05}\textrm{ M}_{\odot}$ and $\cos i = 0.11^{+0.09}_{-0.07}$ \citep{llww05}. However, L\"{o}hmer et al. did not derive a statistically significant constraint on $m_{\rm p}$. A subsequent TOA analysis of the NANOGrav five-year data set \citep{dfg+13} used Markov chain fitting methods and noted issues with the numerical stability of the observed Shapiro timing delay \citep{vv14}. The most recent radio-timing study by \citet{dcl+16} used EPTA data to measure a significant $\dot{x} = 1.07(16)\times10^{-14}$, but did not measure a significant Shapiro delay.

We measured the Shapiro timing delay, $\dot{x} = 1.45(10)\times10^{-14}$ and $\dot{\omega} = -2.8(5)\times10^{-4}\text{ deg yr}^{-1}$ using the NANOGrav nine-year data set for PSR J1640+2224. The $\chi^2$ grids and marginalized PDFs of the Shapiro-delay parameters measured for this MSP are shown in Figure \ref{fig:SDfig2}. Based on the Shapiro timing delay alone, we estimated that $m_{\rm c} = 0.6^{+0.4}_{-0.2}\text{ M}_{\odot}$ and $i = 60(6)$ degrees with the corresponding $m_{\rm p} = 4.4^{+2.9}_{-2.0}\text{ M}_{\odot}$. The highly-significant $\dot{x}$, consistent with the estimate made by \citet{dcl+16} at the $2\sigma$ uncertainty level, is most likely due to a secular change in the inclination of the wide binary system induced by proper motion; the current data set is not sensitive to annual orbital parallax since the annual astrometric parallax was not found to be significant for PSR J1640+2224 \citep{mnf+16}. However, we could not reconcile the $6\sigma$-significant value of $\dot{\omega}$ with the physical mechanisms outlined in Section \ref{subsec:secvar}. In what follows below in this subsection, we explicitly discuss and reject the possibilities that were considered to explain the $\dot{\omega}$ measurement.

The general-relativistic component of $\dot{\omega}$ cannot be the dominant term since our observed value is negative. We also rule out a significant detection of $(\dot{\omega})_{\rm GR}$ since, given the fitted Keplerian elements listed in Table \ref{tab:binarypar}, its predicted value for large assumed component masses is on the order of $10^{-6}\text{ deg yr}^{-1}$. Furthermore, we reject the possibility of this measurement arising from secular orbital variations due to proper motion, since the predicted magnitude of $(\dot{\omega})_{\mu}$ is also on the order of $10^{-6}\text{ deg yr}^{-1}$. 

In principle, a nonzero value of $\dot{\omega}$ can arise from a spin-induced quadrupole term in the companion's gravitational potential due to classical spin-orbit coupling \citep{wex98}; this effect has been observed in pulsar-binary systems with main-sequence companions \citep[e.g.][]{wjm+98}, and can also be observed in pulsar-WD systems in the case where a quadrupole term is induced from rapid rotation of the WD companion. This scenario was first considered in early studies of the relativistic PSR J1141-6545 system by \citet{klm+00}, where they noted that classical spin-orbital coupling would cause a time derivative in the system inclination angle, $di/dt$, that is comparable in order of magnitude to the component of $\dot{\omega}$ due to spin-orbit coupling. We used the $\dot{x}$ measured in the PSR J1640+2224 system, the fact that $\dot{x} = d(a_p\sin i)/dt \approx (a_p\cos i)$ $di/dt$, and the Shapiro-delay estimate of $\sin i$ to compute the time rate of change in the system inclination, and found that $di/dt \sim 10^{-6}\text{ deg yr}^{-1}$. This estimate of $di/dt$ is two orders of magnitude smaller than the observed $\dot{\omega}$, and we therefore reject the significance of classical spin-orbit coupling in our measurement of $\dot{\omega}$ in the PSR J1640+2224 system.

While third-body effects can give rise to measurable perturbations of the pulsar-binary's Keplerian elements \citep[e.g.][]{ras94}, such interactions with another massive component would first be observed as large variations in pulsar-spin period. Our timing solution for PSR J1640+2224 does not show such variations in spin frequency, and so there is no evidence that J1640+2224 is a triple system. Future observations of J1640+2224, along with historical data used by \citet{llww05} and the EPTA data set, will permit for even more stringent estimates of binary-parameter variations evaluated over a larger number of orbits, and ultimately yield a more robust timing solution.

\subsection{PSR J1741+1351}

PSR J1741+1351 is a 3.7-ms pulsar in a 16.3-day orbit that was discovered in a survey of high Galactic latitudes using the Parkes radio telescope \citep{jbo+07}. The Shapiro delay was initially detected in this system by \citet{fjb+06}.

\begin{turnpage}
\begin{figure}
    \centering
    \includegraphics[scale=0.45]{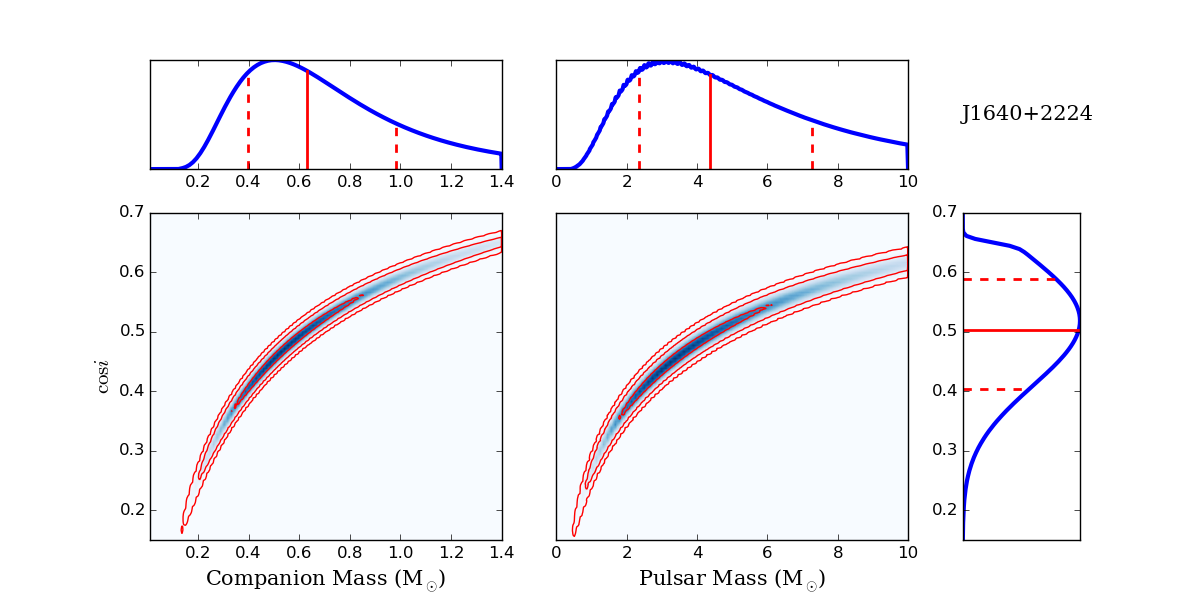}
    \includegraphics[scale=0.45]{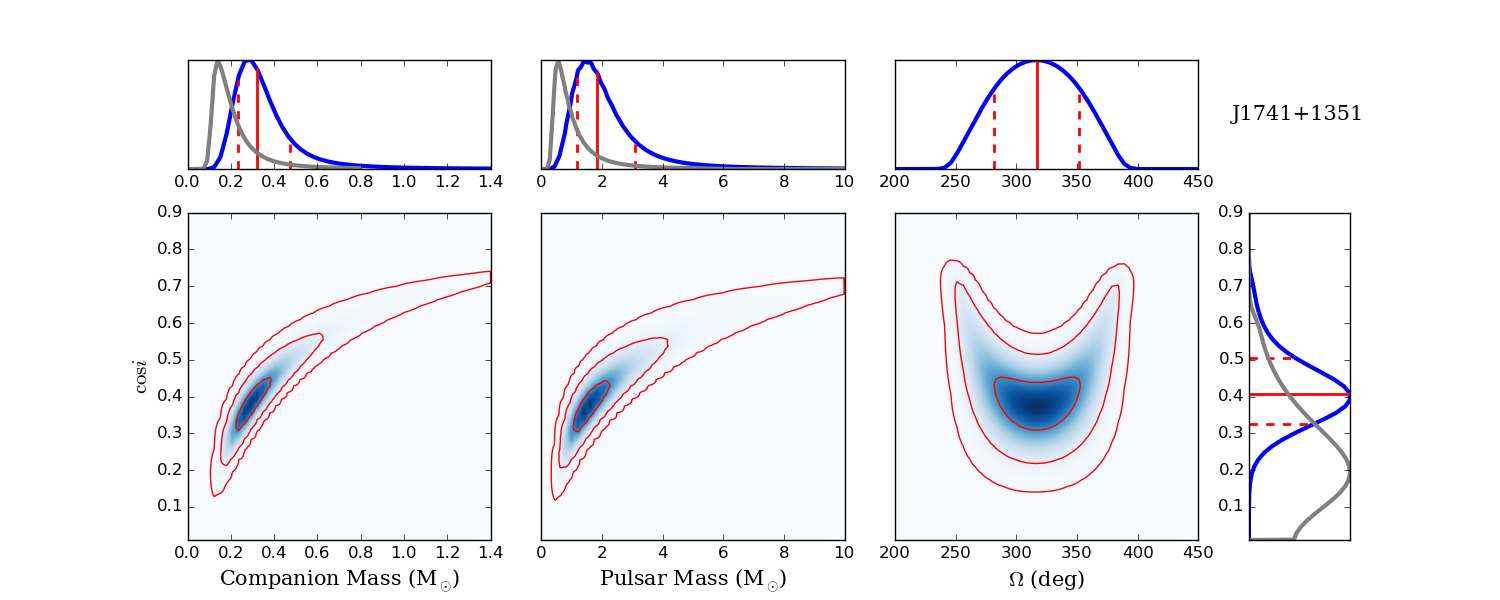}
    \caption{Probability maps and posterior PDFs of the traditional Shapiro-delay parameters for PSR J1640+2224, as well as maps for the Shapiro-delay parameters and $\Omega$ measured for PSR J1741+1351. The inner, middle and outer red contours encapsulate 68.3\%, 95.4\% and 99.7\% of the total probability defined on each two-dimensional map, respectively. In all slimmer panels, the blue solid lines represent posterior PDFs obtained from marginalizing the appropriate two-dimensional map, the vertical red-dashed lines are bounds of the 68.3\% credible interval, and the red-solid line is the median value. Shown for comparison, the grey curves in the slimmer panels of PSR J1741+1351 are marginalized PDFs obtained from computing a separate, two-dimensional $\chi^2$ grid over the ($m_{\rm c}$, $\cos i$) parameters while letting $\dot{x}$ be a free parameter in each TEMPO2 fit.}
    \label{fig:SDfig2}
\end{figure}
\end{turnpage}

We detected the Shapiro timing delay in the NANOGrav nine-year data set for PSR J1741+1351, as well as a highly significant measurement of $\dot{x}$ that we report for the first time. The annual orbital parallax is not significant for this MSP since the annual astrometric parallax was not significantly measured \citep{mnf+16}. As discussed in Section \ref{subsubsec:XDOTconstraints} above, we nonetheless generated a three-dimensional $\chi^2$ grid for different values of the two Shapiro-delay parameters and $\Omega$, in order to constrain the system geometry using both measurements. Figure \ref{fig:SDfig2} shows the $\chi^2$-grid results for PSR J1741+1351 when first generating a three-dimensional, uniform grid in the ($m_{\rm c}$, $\cos i$, $\Omega$) parameters. The two-dimensional ($\cos i$, $\Omega$) probability grid, obtained by marginalizing over $m_{\rm c}$, illustrates a highly non-elliptical covariance between the two parameters. The constrained estimates of the Shapiro-delay parameters are $m_{\rm p} = 1.87^{+1.26}_{-0.69}\text{ M}_{\odot}$, $m_{\rm c} = 0.32^{+0.15}_{-0.09}\text{ M}_{\odot}$, $i = 66^{+5}_{-6}$ degrees, and $\Omega = 317(35)$ degrees.

For comparison, we over-plotted the posterior PDFs obtained from a standard two-dimensional $\chi^2$ grid over the traditional ($m_{\rm c}$, $\cos i$) parameters, while allowing $\dot{x}$ and all other parameters to vary freely in each timing-model fit, as the grey lines in Figure \ref{fig:SDfig2}. There are clear and significant differences between the posterior PDFs, which strongly suggest correlation between $\dot{x}$ and one or both of the Shapiro delay parameters. The three-dimensional $\chi^2$-grid results indicate that explicit modeling of the highly-significant kinematic term reduces correlation between the Shapiro-delay parameters and $\dot{x}$, and produces more sensible posterior PDFs of the component masses and system inclination that are consistent with initial results presented by \citet{fjb+06}. 

\subsection{PSR B1855+09}

PSR B1855+09 is a 5.4-ms pulsar in a 12.3-day orbit with a WD companion, and is also one of the earliest MSP discoveries made using the Arecibo Observatory \citep{srs+86}. This MSP-binary system was the first to yield a significant measurement of the Shapiro timing delay from pulsar-timing measurements \citep{rt91a}. The most recent long-term radio timing study determined the pulsar mass to lie within the range $1.4 < m_{\rm p} < 1.8\text{ M}_{\odot}$ \citep[95\% confidence;][]{nss04b}. Optical follow-up observations of the companion yielded a WD-cooling timescale of $\sim$10 Gyr, which is twice as long as the characteristic age of the MSP \citep{vbkk00}. 

\begin{figure}
    \centering
    \includegraphics[scale=0.45]{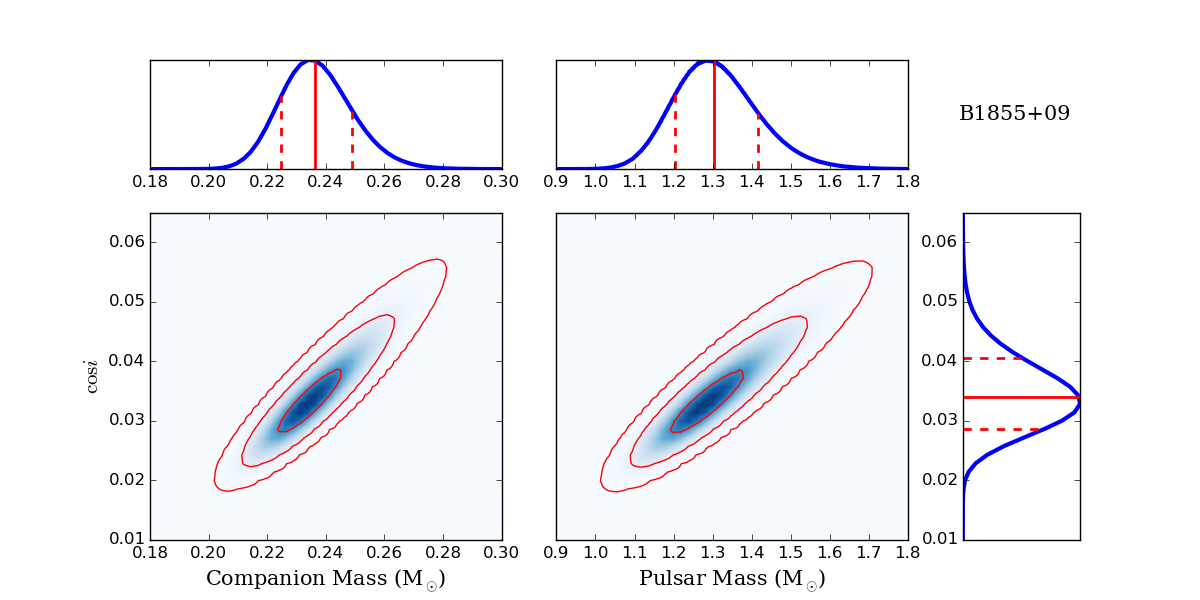}
    \includegraphics[scale=0.45]{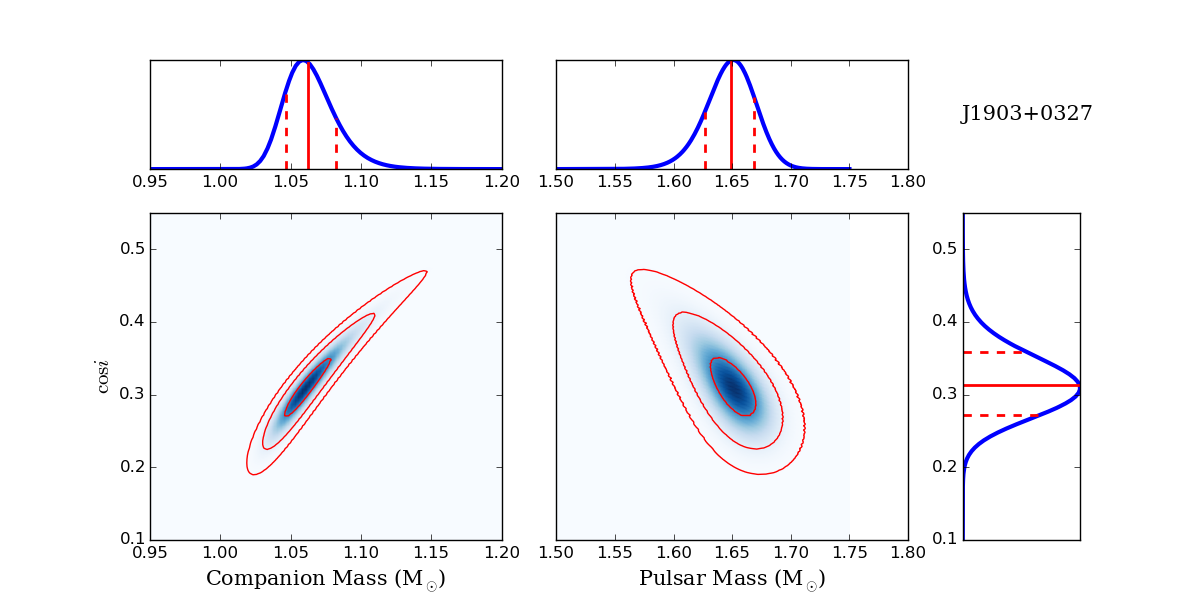}
    \includegraphics[scale=0.45]{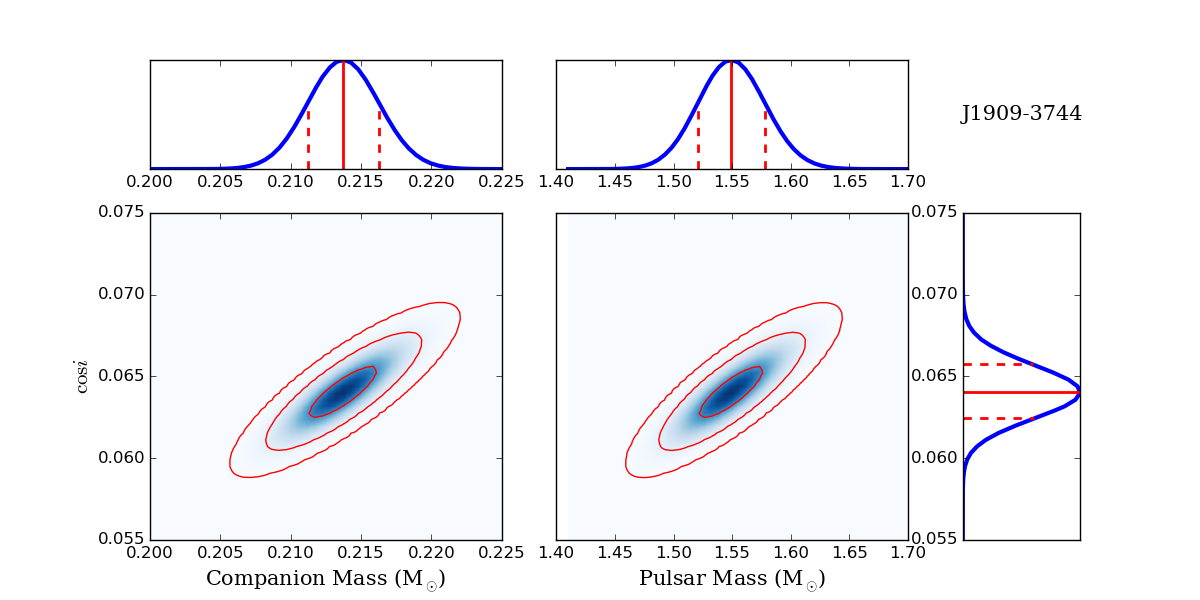}
    \caption{Probability maps and posterior PDFs of the Shapiro-delay parameters measured for PSRs B1855+09, J1903+0327, and J1909$-$3744. The maps and PDFs for J1903+0327 were constrained assuming that the observed $\dot{\omega}$ is due to GR (see Section \ref{subsec:J1903}). The inner, middle and outer red contours encapsulate 68.3\%, 95.4\% and 99.7\% of the total probability defined on each two-dimensional map, respectively. In the slimmer panels, the blue solid lines represent posterior PDFs obtained from marginalizing the appropriate two-dimensional map, the vertical red-dashed lines are bounds of the 68.3\% credible interval, and the red-solid line is the median value.}
    \label{fig:SDfig3}
\end{figure}

We made a highly significant measurement of the Shapiro timing delay when using the NANOGrav nine-year data set for PSR B1855+09. The $\chi^2$ grids and marginalized PDFs for PSR B1855+09 are shown in Figure \ref{fig:SDfig3}.  Our estimates of the component masses and inclination angle --  $m_{\rm p} = 1.30^{+0.11}_{-0.10}\text{ M}_{\odot}$, $m_{\rm c} = 0.236^{+0.013}_{-0.011}\text{ M}_{\odot}$, and $i = 88.0^{+0.3}_{-0.4}$ degrees -- are consistent with, and more precise than, those previously made by \citet{ktr94}, \citet{nss04b} and \citet{rhc+16}. 

\subsection{PSR J1903+0327}
\label{subsec:J1903}

PSR J1903+0327 is a 2.1-ms pulsar in an eccentric, 95-day orbit with a main-sequence companion \citep{crl+08}. This binary system, located within the Galactic disk, posed a significant challenge to the standard view of MSP formation since tidal interactions are expected to produce low-eccentricity orbits with WD companions, as is observed for all other disk MSP-binary systems. \citet{fbw+11} performed the most recent pulsar-timing analysis of PSR J1903+0327 and argued that both binary components were once members of a progenitor triple system where the main-sequence companion was in an outer orbit about an inner MSP-WD binary; this system was subsequently disrupted and produced the binary currently observed, either by a chaotic third-body interaction or full dissipation of the inner WD companion. They combined their Shapiro-delay measurement for this system with a significant measurement of $\dot{\omega}$, which they argue is due to GR, to determine the component masses and inclination with high precision: $m_{\rm p} = 1.667(21)\textrm{ M}_{\odot}$; $m_{\rm c} = 1.029(8)\text{ M}_{\odot}$; and $77.47(15)$ degrees (all 99.7\% confidence). Freire et al. also measured an $\dot{x} = 0.020(3)\times10^{-12}$ that they attributed to proper-motion bias. A recent optical analysis of radial-velocity measurements estimated the mass ratio of this system to be $q = m_{\rm p}/m_{\rm c} = 1.56(15)$ \citep[68.3\% confidence;][]{ksf+12}, consistent with the radio-timing estimate of $q$ = 1.62(3) made by Freire et al. 

We also independently measure a significant $\dot{\omega} = 2.410(13)\times10^{-4}\text{ deg yr}^{-1}$ in the PSR J1903+0327 system, as well as the Shapiro timing delay indicated by the significance of $h_3$ listed in Table \ref{tab:postKeppar}. We do not measure a significant $\dot{x}$. The observed $\dot{\omega}$ from our data set is consistent with the measurement made by \citet{fbw+11}, and so we used the methodology discussed in Section \ref{subsubsec:OMDOTconstraints} to constrain the Shapiro-delay parameters assuming that GR describes the observed periastron shift. The constrained $\chi^2$ grids for PSR J1903+0327 are shown in Figure \ref{fig:SDfig3}. From these grids, we estimated the component masses and inclination to be: $m_{\rm p} = 1.65(2)\text{ M}_{\odot}$; $m_{\rm c} = 1.06(2)\text{ M}_{\odot}$; and $i = 72^{+2}_{-3}\text{ deg yr}^{-1}$. The estimate of $m_{\rm p}$ agrees with the Freire et al. measurement at the 68.3\% credibility level, while $m_{\rm c}$ and $i$ are consistent at about the 95.4\% credibility level. We do not adjust the uncertainty in our measurement of $\dot{\omega}$ for the maximum uncertainty in $(\dot{\omega})_{\mu}$, which Freire et al. do when deriving their estimates. Our derived estimate of $q = 1.56(3)$ also agrees with the optical measurement and Freire et al. estimate mentioned above.

\subsection{PSR J1909$-$3744}

PSR J1909-3744 is a 2.9-ms pulsar in a 1.5-day orbit with a WD companion \citep{jhb+05}. The Shapiro timing delay has previously been observed in this system with high precision, leading to the first precise mass measurement for an MSP \citep{jhb+05, hbo06, ver09}. Two recent, independent TOA analyses of this pulsar were performed by \citet{rhc+16} and \citet{dcl+16}. Reardon et al. used the PPTA data set and reported significant Shapiro-delay parameters, apparent orbital decay, and geometric variations the PSR J1909-3744 system with the following measured and derived results: $m_{\rm p} = 1.47(3)\text{ M}_{\odot}$; $m_{\rm c} = 0.2067(19)\text{ M}_{\odot}$; $i = 93.52(9)^{\circ}$; and $\dot{P}_{\rm b} = 0.503(6)\times10^{-12}$. Desvignes et al. analyzed the EPTA data set and also reported estimates of the Shapiro-delay parameters, apparent orbital decay, and geometric variations: $m_{\rm p} = 1.54(3)\text{ M}_{\odot}$; $m_{\rm c} = 0.213(2)\text{ M}_{\odot}$; $\sin i = 0.99771(13)$; and $\dot{P}_{\rm b} = 0.503(5)\times10^{-12}$.

We independently measure both Shapiro-delay parameters and $\dot{P}_{\rm b}$ with high significance when using the NANOGrav nine-year data set. We also make a marginal detection of $\dot{x} = -4.4(1.6)\times10^{-16}$ when incorporating it as a free parameter, but it does not pass the F-test criterion. 

The component masses that we derived from the probability maps for J1909$-$3744 shown in Figure \ref{fig:SDfig3}, $m_{\rm p} = 1.55(3)\text{ M}_{\odot}$ and $m_{\rm c} = 0.214(3)\text{ M}_{\odot}$, agree with the estimates made by \citet{rhc+16} and \citet{dcl+16}. Our estimate of $i = 86.33(10)$ degrees possesses a sign ambiguity in $\cos i$, so $i = 93.67(10)$ is an allowed solution for our analysis; the latter estimate agrees with the Reardon et al. and Desvignes et al. measurement. 

Given our measurements of the Keplerian and Shapiro-delay parameters,  the expected orbital decay in this system from quadrupole gravitational-wave emission is $(\dot{P}_{\rm b})_{\rm GR} = -0.00294 \times 10^{-12}$, which is significantly less than our measurement of $\dot{P}_{\rm b}$. This low estimate of $(\dot{P}_{\rm b})_{\rm GR}$ implies that $\dot{P}_{\rm b} = 0.509(9)\times10^{-12} \approx (\dot{P}_{\rm b})_{D}$, which agrees with the measurement and assessment made by \citet{rhc+16} and \citet{dcl+16}. We therefore attribute the apparent orbital decay in PSR J1909$-$3744 system to biases from significant acceleration between the MSP-binary and SSB reference frames. \citet{mnf+16} used our $\dot{P}_{\rm b}$ measurement to find the distance to PSR J1909-3744 to be 1.11(2)~kpc, in agreement with their timing-parallax distance of $1.07^{+0.04}_{-0.03}$ kpc.

\subsection{PSR J1918$-$0642}

PSR J1918$-$0642 is a 7.6-ms pulsar in a 10.9-day orbit with a likely WD companion that was discovered by \citet{eb01} in a multi-beam survey of intermediate Galactic latitudes using the Parkes Radio Telescope. An optical search for the companion of PSR J1918$-$0642 was unsuccessful \citep{vbjj05}, requiring that the apparent R-band magnitude of the WD be $R > 24$. A long-term timing study of this MSP was carried out by \citet{jsb+10} using the Westerbork, Nan\c{c}ay and Jodrell Bank radio observatories at 1400 MHz for a combined timespan of 7.4 years. While only Keplerian parameters were measured, \citet{jsb+10} combined their distance estimate to PSR J1918$-$0642 -- based on their dispersion-measure estimate for this pulsar and the \citet{cl01} electron-density model for the Galaxy -- with the $R > 24$ limit, and the assumption that the white-dwarf cooling and pulsar spin-down are coeval, to further constrain the companion to be a He or CO white dwarf with a thin hydrogen atmosphere. They used the mass function of the system, as well as an assumed $m_{\rm p} = 1.35\textrm{ M}_{\odot}$, to compute a minimum companion mass of $m_{\rm c, min} = 0.24\textrm{ M}_{\odot}$. A recent radio-timing analysis by \citet{dcl+16} used the EPTA data to measure the Shapiro delay in this system, with $m_{\rm p} = 1.3^{+0.6}_{-0.4}\text{ M}_{\odot}$, $m_{\rm c} = 0.23(7)\text{ M}_{\odot}$, and $\cos i = 0.09^{+0.05}_{-0.04}$.

\begin{figure}
    \centering
    \includegraphics[scale=0.45]{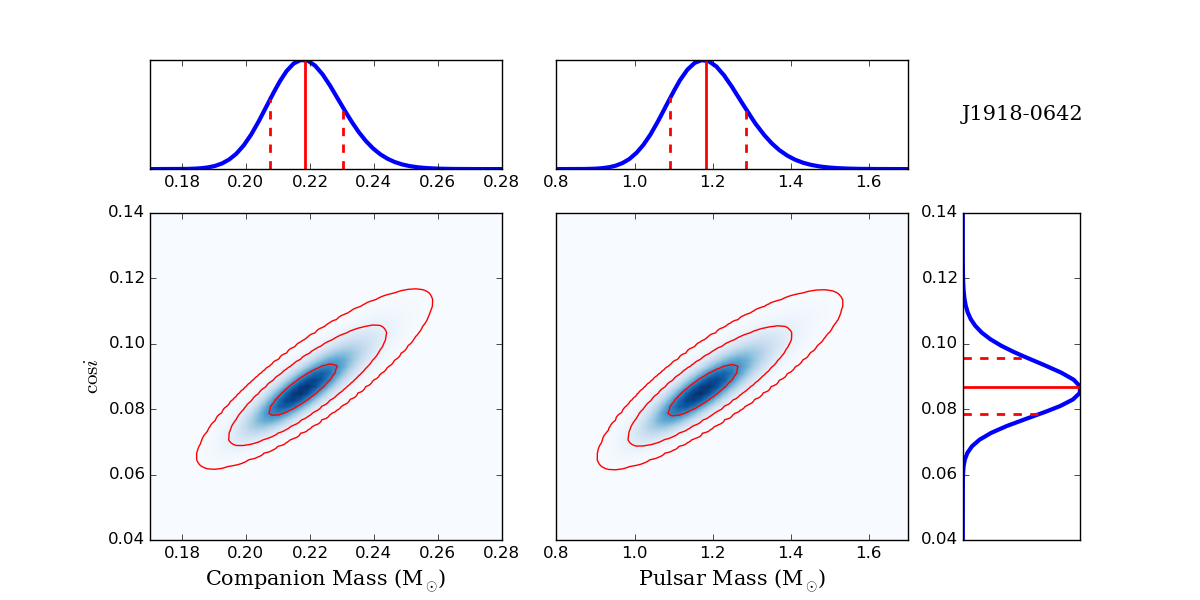}
    \includegraphics[scale=0.45]{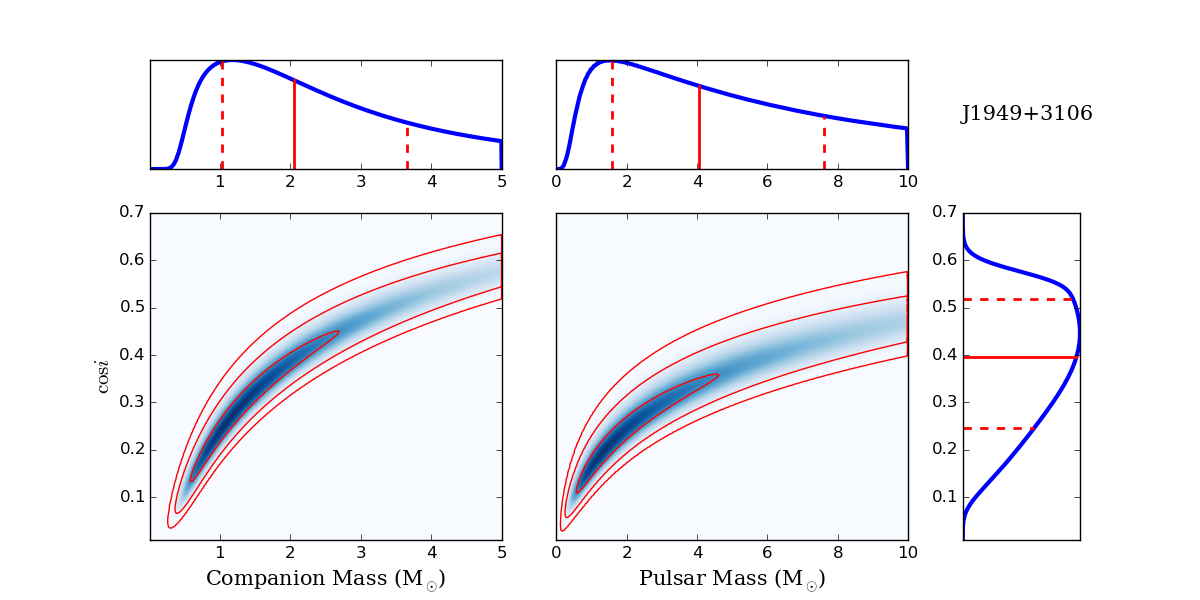}
    \includegraphics[scale=0.45]{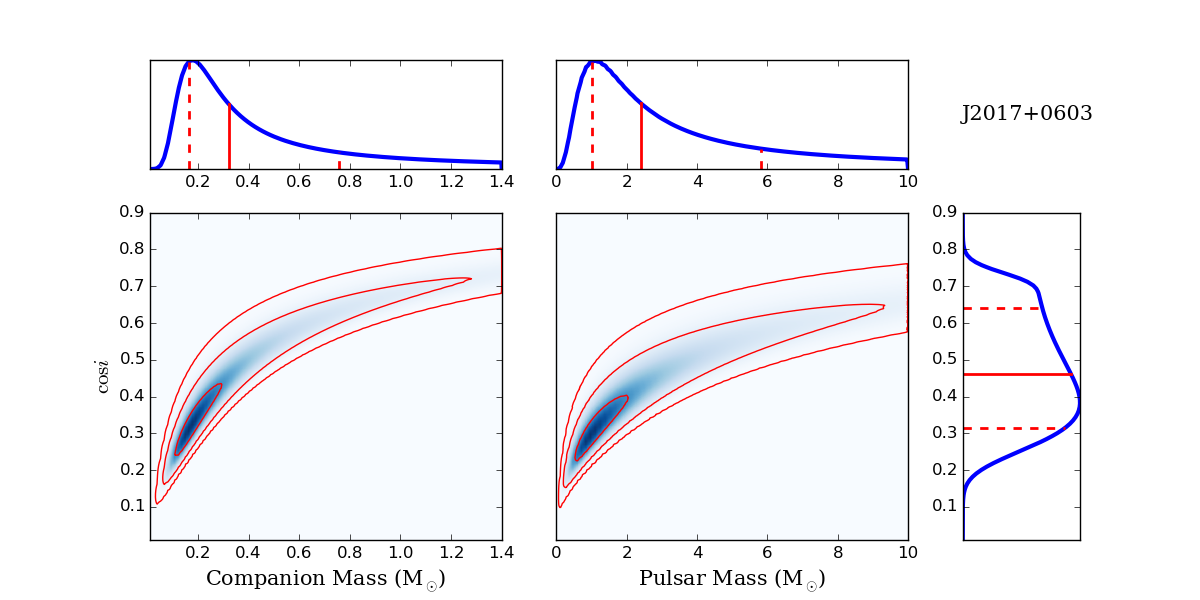}
    \caption{Probability maps and posterior PDFs of the Shapiro-delay parameters measured for PSRs J1918$-$0642, J1949+3106, and J2017+0603. The inner, middle and outer red contours encapsulate 68.3\%, 95.4\% and 99.7\% of the total probability defined on each two-dimensional map, respectively. In the slimmer panels, the blue solid lines represent posterior PDFs obtained from marginalizing the appropriate two-dimensional map, the vertical red-dashed lines are bounds of the 68.3\% credible interval, and the red-solid line is the median value.}
    \label{fig:SDfig4}
\end{figure}

We measured a highly-significant Shapiro timing delay in the PSR J1918$-$0642 binary system using the NANOGrav nine-year data set. The probability maps computed from $\chi^2$ grids for PSR J1918$-$0642 are shown in Figure \ref{fig:SDfig4}. The significance of $h_3$ in the PSR J1918$-$0642 system exceeds 27$\sigma$, a factor of $\sim 4$ better than the $h_3$ estimate made by \citet{dcl+16} when using their EPTA data set. Our precise measurements of the WD mass and inclination from the Shapiro timing delay are $m_{\rm c} = 0.219^{+0.012}_{-0.011}\textrm{ M}_{\odot}$ and $i = 85.0(5)$ degrees regardless of choice in the parameterization of $\Delta_{\rm S}$. The derived estimate of the pulsar mass is the first precise estimate for this system, and is suggestive of a low-mass neutron star: $m_{\rm p} = 1.18^{+0.10}_{-0.09}\textrm{ M}_{\odot}$.

\subsection{PSR J1949$+$3106}

PSR J1949+3106 is a 13.1-ms pulsar in a 1.9-day orbit with a massive companion that was discovered by the ongoing PALFA survey of the Galactic plane using the Arecibo telescope \citep{dfc+12}. The initial radio-timing study by Deneva et al. used TOAs collected with the Arecibo, Green Bank, Nan\c{c}ay and Jodrell Bank telescopes over a four-year period to make a significant detection of the Shapiro timing delay in this system. They reported significant measurements of the orthometric parameters, $h_3 = 2.4(1)\text{ }\mu\text{s}$ and $\varsigma = 0.84(2)$, as well as derived estimates of component masses and system inclination: $m_{\rm p} = 1.47^{+0.43}_{-0.31}\text{ M}_{\odot}$; $m_{\rm c} = 0.85^{+0.14}_{-0.11}\text{ M}_{\odot}$; and $i = 79.9^{+1.6}_{-1.9}$ degrees.

We independently measured a Shapiro timing delay in the PSR J1949+3106 using the NANOGrav nine-year data set. The probability maps computed from $\chi^2$ grids for PSR J1949+3106 are shown in Figure \ref{fig:SDfig4}; we set $m_{\rm c, max} = 5\textrm{ M}_{\odot}$ when computing the $\chi^2$ grids since the peak-probability value is nearly equal to our usual upper limit of $m_{\rm c, max} = 1.4\textrm{ M}_\odot$. Our measurements of the orthometric parameters, $h_3$ = 2.5(5) $\mu$s and $\varsigma = 0.77(10)$, are consistent with those made by \citep{dfc+12} at the 68.3\% credibility level. The uncertainties in our measurements are comparatively larger due to the shorter time span of our data set and, therefore, less TOA coverage across the orbit. Our derived estimates of the component masses and inclination are subsequently much less stringent than those made by Deneva et al.: $m_{\rm p} = 4.0^{+3.6}_{-2.5}\text{ M}_{\odot}$; $m_{\rm c} = 2.1^{+1.6}_{-1.0}\text{ M}_{\odot}$; and $i = 67^{+9}_{-8}$ degrees.

\subsection{PSR J2017+0603}
\label{subsec:J2017}

PSR J2017+0603 is a 2.9-ms pulsar in a 2.2-day orbit that was initially found using the {\it Fermi} Large Area Telescope (LAT) as a gamma-ray source with no known associations; radio pulsations were discovered and subsequently timed from this source using the Nancay Radio Telescope and Jodrell Bank Observatory for nearly two years by \citet{cgj+11}. They used the mass function of the PSR J2017+0603 system, along with an assumed $m_{\rm p} = 1.35\text{ M}_{\odot}$, to compute a minimum companion mass of $m_{\rm c, min} = 0.18\text{ M}_{\odot}$.

For the first time, we detect a Shapiro timing delay in the PSR J2017+0603 system using the NANOGrav nine-year data set, with $m_{\rm c} = 0.32^{+0.44}_{-0.16}\text{ M}_{\odot}$ and $i = 62^{+9}_{-12}$ degrees. The probability maps computed from $\chi^2$ grids for PSR J2017+0603 are shown in Figure \ref{fig:SDfig4}. The observed Shapiro delay in this system is currently weak since the marginalized, one-dimensional PDF of $m_{\rm p} = 2.4^{+3.4}_{-1.4}\text{ M}_{\odot}$ extends to large values of the neutron-star mass. However, we were able to make a significant detection using a comparatively small, 1.7-yr data set that includes targeted observations at select orbital phases discussed in Section \ref{sec:obs}; our measurement will improve with the inclusion of future TOAs collected at different points in the orbit. 

\subsection{PSR J2043+1711}
\label{subsec:J2043}

PSR J2043+1711 is a 2.4-ms pulsar in a 1.5-day orbit that was initially found using the {\it Fermi} LAT as a gamma-ray source with no previously known associations. The radio counterpart was discovered using the Nancay and Green Bank Telescopes; the Shapiro delay was detected in this MSP-binary system using a timing model derived from TOAs collected with the Nancay, Westerbork and Arecibo observatories over a three-year period \citep{gfc+12}. At the time of the initial study performed by Guillemot et al., the Shapiro timing delay was not significant enough to yield statistically meaningful estimates of the component masses and inclination angle. They placed limits on the companion mass by assuming the validity of the $m_{\rm c}$-$P_{\rm b}$ relation, and derived a preferred range of $0.20 < m_{\rm c} < 0.22\text{ M}_{\odot}$; with this constraint, Guillemot et al. found the pulsar mass and inclination to be $1.7 < m_{\rm p} < 2.0 \textrm{ M}_{\odot}$ and $i = 81.3(1.0)$ degrees, respectively. 

\begin{figure}
    \centering
    \includegraphics[scale=0.45]{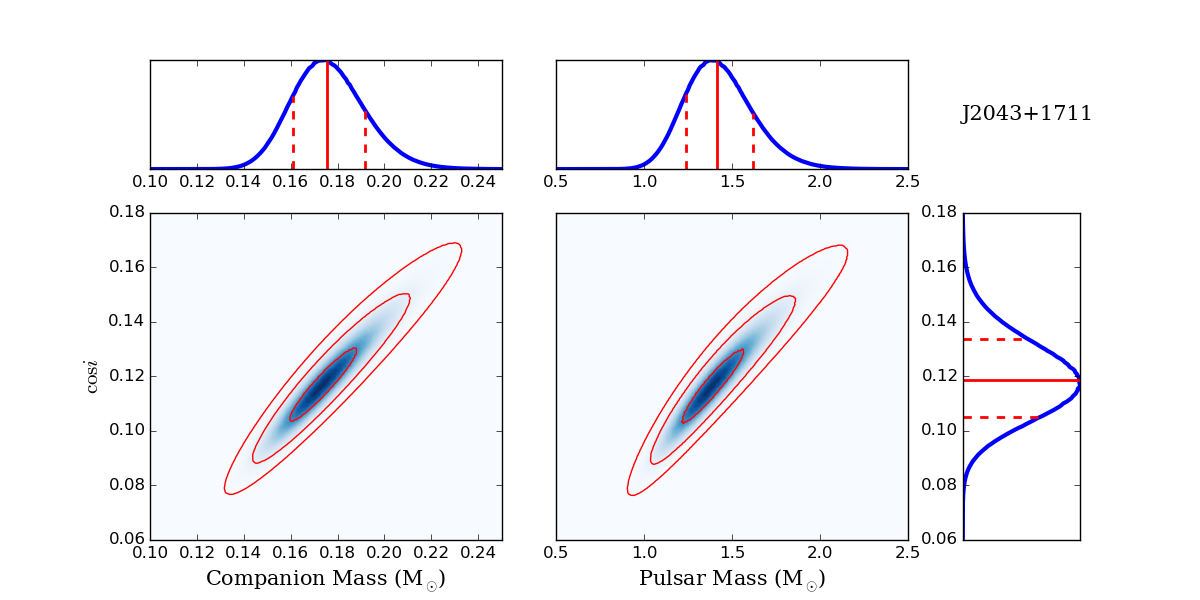}
    \includegraphics[scale=0.45]{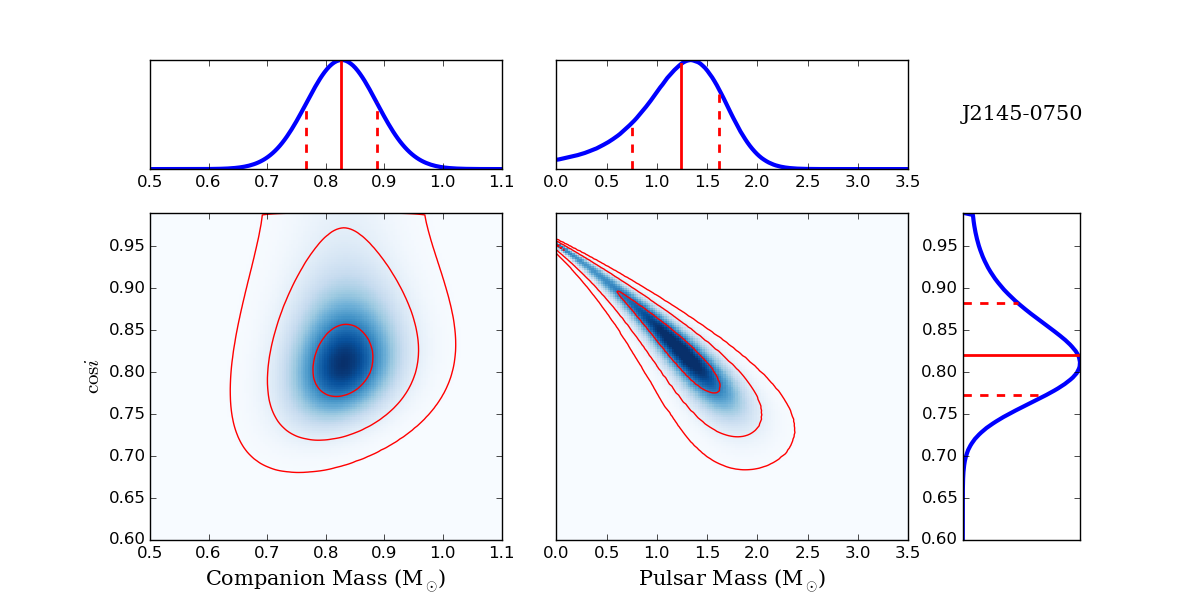}
    \includegraphics[scale=0.45]{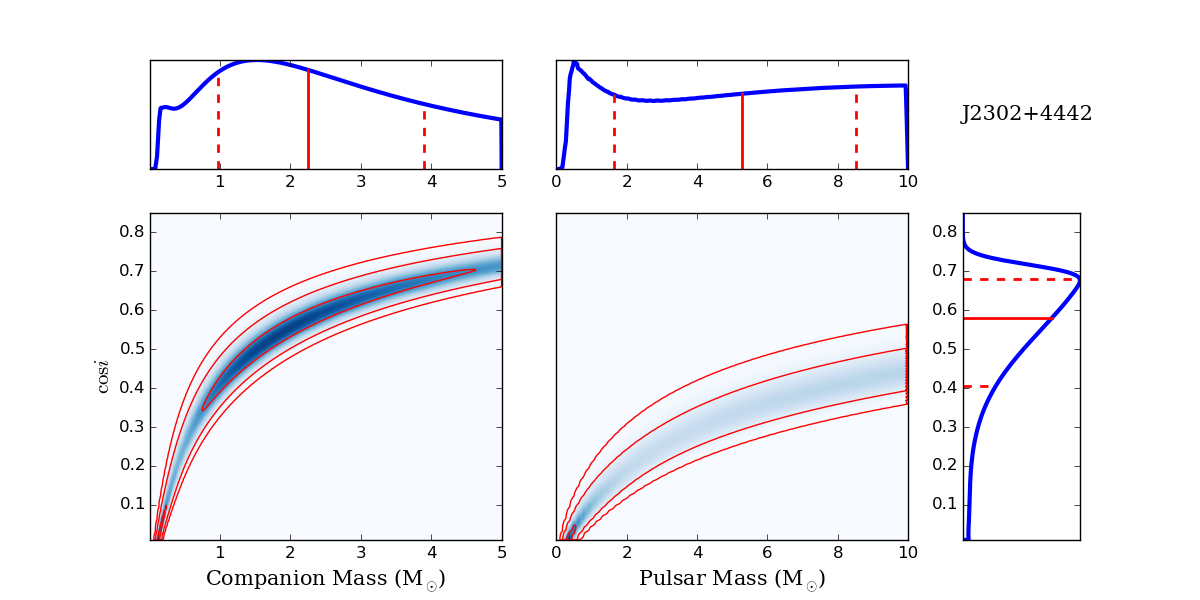}
    \caption{Probability maps and posterior PDFs of the Shapiro-delay parameters measured for PSR J2043+1711, J2302+4442, and J2145$-$0750. The inner, middle and outer red contours encapsulate 68.3\%, 95.4\% and 99.7\% of the total probability defined on each two-dimensional map, respectively. In the slimmer panels, the blue solid lines represent posterior PDFs obtained from marginalizing the appropriate two-dimensional map, the vertical red-dashed lines are bounds of the 68.3\% credible interval, and the red-solid line is the median value.}
    \label{fig:SDfig5}
\end{figure}

The NANOGrav nine-year data set on PSR J2043+1711, which includes the targeted Shapiro-delay observations discussed in Section \ref{sec:obs}, yields a significantly improved measurement of the component masses and system inclination as shown in Table \ref{tab:SDresults}; the impact of the targeted observations on the significance of $\Delta_{\rm S}$ in the PSR J2043+1711 system was discussed by \citet{pen15}. The probability maps computed from $\chi^2$ grids for PSR J2043+1711 are shown in Figure \ref{fig:SDfig5}. Our improved measurements of $m_{\rm c} = 0.175^{+0.016}_{-0.015}\text{ M}_{\odot}$ and $i = 83.2^{+0.8}_{-0.9}$ degrees are consistent with the initial estimates made by \citet{gfc+12}, though $m_{\rm c}$ is moderately lower than the range determined from the $m_{\rm c}$-$P_{\rm b}$ relation. Our derived $m_{\rm p} = 1.41^{+0.21}_{-0.18}\textrm{ M}_{\odot}$ is therefore slightly below the $m_{\rm p}$ range determined by Guillemot et al. when assuming the validity of the $m_{\rm c}-P_{\rm b}$ relation. 

\subsection{PSR J2145$-$0750}
\label{subsec:J2145}

PSR J2145$-$0750 is a 16-ms pulsar in a 6.8-day orbit with a white-dwarf companion and was discovered in a Parkes Telescope survey \citep{bhl+94}. Both \citet{pk94} and \citet{vdh94} argued that the J2145$-$0750 system likely experienced unstable mass transfer from ``common-envelope" evolution, where the pulsar gradually expelled the outer layers of the donor, in order to explain its unusually long pulsar-spin period and massive companion compared to other binary-MSP systems. Early optical observations of the WD companion noted the difficulty in obtaining accurate photometry due to the use of a dispersion-based distance estimate and the presence of a coincident field star \citep{lcf+95}. However, a recent study performed by \citet{dvk+16} combined improved optical imaging with a precise VLBI distance of $d = 613^{+16}_{-14}$ pc to estimate a companion mass of $m_{\rm c} \approx 0.85\textrm{ M}_{\odot}$. Deller et al. also detected the orbital reflex motion of J2145$-$0750 through their VLBI measurements, and inferred estimates of $i = 21^{+7}_{-4}$ degrees and $\Omega = 230(12)$ degrees.\footnote{\citet{dvk+16} report their estimate of $\Omega$ using a convention that measures $\Omega$ from celestial East through North. This convention is inconsistent with the North-through-East convention we use in this work. We report their estimate of $\Omega$ relative to our convention.}

We measured $\dot{x} = 0.0098(19)\times10^{-12}$, consistent with estimates made by \citet{rhc+16}. Our estimate of $h_3 = 0.10(5)\textrm{ }\mu$s does not pass the $h_3$-significance test, and so we do not formally measure a significant Shapiro timing delay from the radio-timing data alone. However, we used the estimate of $m_{\rm c} = 0.83^{+0.06}_{-0.06}\textrm{ M}_{\odot}$ made by \citet{dvk+16} as a prior distribution when computing the posterior maps for PSR J2145$-$0750. The resulting constraints on $\cos i$ and $m_{\rm p}$ are shown in Figure \ref{fig:SDfig5}, which yield $m_{\rm p} = 1.3^{+0.4}_{-0.5}\textrm{ M}_{\odot}$ and $i = 34^{+5}_{-7}$ degrees, and are consistent with estimates made Deller et al.

\subsection{PSR J2302+4442}
\label{subsec:J2302}

PSR J2302+4442 is a 5.2-ms pulsar in a 126-day orbit that, along with PSR J2017+0603  (Section \ref{subsec:J2017}) was initially found using the {\it Fermi} LAT as a gamma-ray source with no known associations and observed in the radio using the Nan\c{c}ay Radio Telescope and Jodrell Bank Observatory for nearly two years by \citet{cgj+11}. They used the mass function of the PSR J2302+4442 system, along with an assumed $m_{\rm p} = 1.35\text{ M}_{\odot}$, to compute a minimum companion mass of $m_{\rm c, min} = 0.3\text{ M}_{\odot}$.

For the first time, we tentatively detect a Shapiro timing delay in the PSR J2302$+$4442 system using the NANOGrav nine-year data set. The probability maps computed from $\chi^2$ grids for PSR J2302+4442 are shown in Figure \ref{fig:SDfig5}. Due to the weak detection of $\Delta_{\rm S}$ and large correlation between $r$ and $s$, the timing solution published by \citet{abb+15b} used a fixed value of $m_{\rm c} = 0.355\text{ M}_{\odot}$ that was computed from the $m_{\rm c}$-$P_{\rm b}$ relation when fitting for all other timing parameters, including the Shapiro $s$ parameter. In this study, we developed timing solutions using both the traditional and orthometric parameterizations of $\Delta_{\rm S}$ that allowed both PK parameters to be fitted for. The value of $h_3$ in the PSR J2302+4442 system exceeds $5\sigma$ and therefore passes the $h_3$ significance test for detection of $\Delta_{\rm S}$.

Our estimates of the companion mass and inclination are $m_{\rm c} = 2.3^{+1.7}_{-1.3}\text{ M}_{\odot}$ and $i = 54^{+12}_{-7}$ degrees, and the corresponding pulsar mass is $m_{\rm p} = 5.3^{+3.2}_{-3.6}\text{ M}_{\odot}$. We computed $\chi^2$ grids with $m_{\rm c,max} = 5\textrm{ M}_{\odot}$ since the peak-probability value of $m_{\rm c}$ exceeds the usual upper limit of $m_{\rm c, max} = 1.4\textrm{ M}_\odot$. While the posterior PDFs of the component masses span a large range of mass values, the significant estimates of $s$ and $\varsigma$ indicate a measurable constraint on the system inclination. The measurement of $\Delta_{\rm S}$ will improve in significance over time since the current data set for PSR J2302+4442 only spans about 1.7 years -- or $\sim$5 orbits, given the long $P_{\rm b}$ of this MSP-binary system -- and so a very small fraction of the Shapiro-delay signal has been sampled. Furthermore, given the large orbit and modest inclination, we expect to see a measurable secular variation in $x$ within the next few years. 

\subsection{PSR J2317+1439}
\label{subsec:J2317}

PSR J2317+1439 is a 3.4-ms pulsar in a 2.5-day orbit that was discovered in a survey of high Galactic latitudes using the Arecibo Obsveratory and possesses one of the smallest eccentricities known \citep{cnt93,cnt96,hlk+04}. The most recent radio-timing analysis of PSR J2317+1439 performed by \citet{dcl+16} did not yield any secular variations in orbital parameters or a significant measurement of the Shapiro timing delay when using their 17.3-yr EPTA data set. However, a Bayesian-timing analysis performed by \citet{vv14} used the NANOGrav five-year data set \citep{dfg+13} to measure several secular variations in the binary parameters: $\dot{P}_{\rm b} = 6.4(9)\times10^{-12}$; $\dot{\eta} = -2(4)\times10^{-15}$; and $\dot{\kappa} = 2.0(7)\times10^{-14}$. Vigeland and Vallisneri noted that many of the posterior distributions for binary parameters of J2317+1439 changed slightly when using different priors for the astrometric timing parallax.

The original NANOGrav nine-year timing model for PSR J2317+1439 contains parameters that describe secular variations in $x$ and the Laplace-Lagrange eccentricity parameters, with $\dot{\eta} = 5.0(9)\times10^{-15}\text{ s}^{-1}$, all of which pass the F-test criterion. We found that $\dot{P}_{\rm b}$ did not pass the F-test, so it was not fitted in the original NANOGrav nine-year timing solution. Moreover, both the F-test and the $h_3$-significance test indicated that the Shapiro delay was not significant, and so we also did not initially incorporate the Shapiro-delay parameters.

Despite the statistical significance of $\dot{\eta}$, we do not believe that the PSR J2317+1439 system is experiencing physical processes that produce a changing eccentricity. For instance, if mass transfer between components were currently taking place, we would expect to observe a spin-up phase; instead, we observe seemingly ``normal" spin-down properties and stable rotation that is typical of MSPs. The presence of a third massive body in a bound, hierarchical orbit about the pulsar-companion binary system would induce higher-order derivatives in spin frequency as well as additional third-body effects on the shape, size and period of the inner binary \citep[e.g.][]{jr97}, most of which we do not see in the NANOGrav nine-year data set. Finally, the timescale for the observed change in $\eta$ is estimated to be $\eta/\dot{\eta} \approx 0.7$ years, which is implausibly short.

Because the observed $\dot{\eta}$ is physically implausible, and because covariances between it and several other parameters distort the timing solution, we chose to hold both $\dot{\eta}$ and $\dot{\kappa}$ fixed to a value of zero (i.e. no change in the eccentricity parameters of the system) while re-fitting the nine-year timing model. In this case, we found that the significance of $h_3$ exceeded $3\sigma$ and therefore included the Shapiro-delay parameters. We found that $\dot{x}$ did not pass the F-test, and so did not fit for it in our modified solution. The new timing model for PSR J2317+1439 fits the data well (reduced $\chi^2$ = 1.0053 for 2531 degrees of freedom), though the original model published by \citet{abb+15b} that fits for $\dot{\eta}$ and $\dot{\kappa}$ better fits the TOA data (reduced $\chi^2$ = 0.9966 for 2531 degrees of freedom).  

\begin{figure}
    \centering
    \includegraphics[scale=0.45]{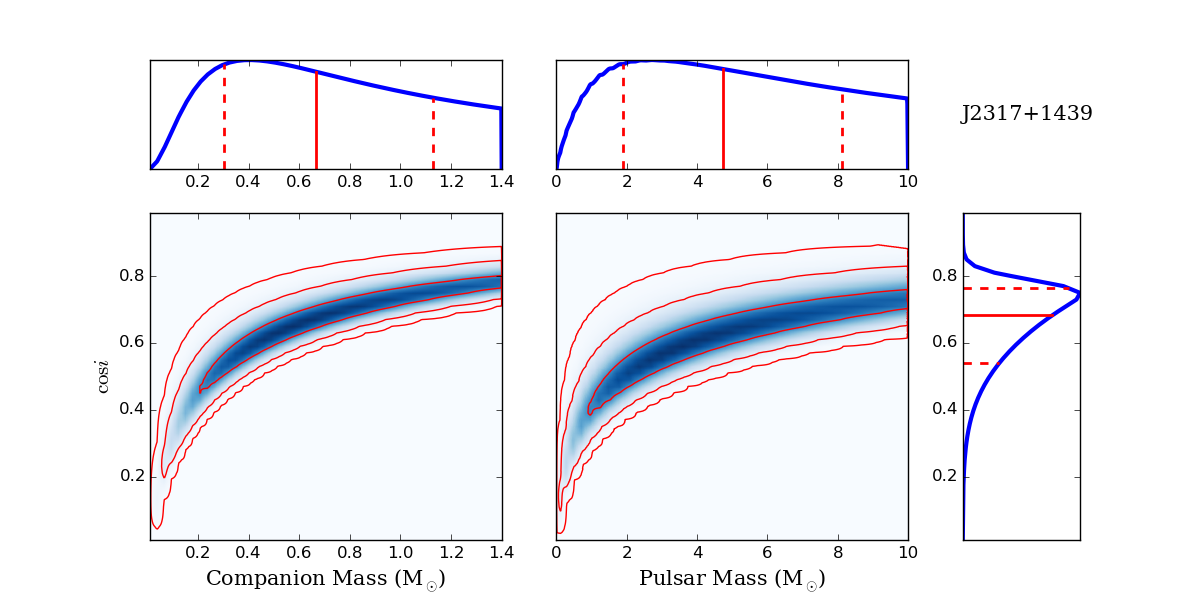}
    \caption{Probability maps and posterior PDFs of the Shapiro-delay parameters measured for PSR J2317+1439. The inner, middle and outer red contours encapsulate 68.3\%, 95.4\% and 99.7\% of the total probability defined on each two-dimensional map, respectively. In the slimmer panels, the blue solid lines represent posterior PDFs obtained from marginalizing the appropriate two-dimensional map, the vertical red-dashed lines are bounds of the 68.3\% credible interval, and the red-solid line is the median value.}
    \label{fig:SDfig6}
\end{figure}

We generated two-dimensional $\chi^2$ grids for the traditional and orthometric Shapiro-delay parameters. The probability maps and the marginalized PDFs of the component masses and system inclination are shown in Figure \ref{fig:SDfig6}. Given the new binary timing model of PSR J2317+1439, we have made a weak detection of the Shapiro timing delay in this system since the two-dimensional probability density extends to large $m_c$ for low inclinations, and so the system inclination angle is not as well constrained as for the other stronger detections. Our current estimates of the component masses and inclinations are $m_{\rm p} = 4.7^{+3.4}_{-2.8}\text{ M}_{\odot}$, $m_{\rm c} = 0.7^{+0.5}_{-0.4}\text{ M}_{\odot}$, and $i = 47^{+10}_{-7}$ degrees.

\section{Conclusions \& Summary}
\label{sec:conclusions}

\begin{figure}
    \centering
    \includegraphics[scale=0.6]{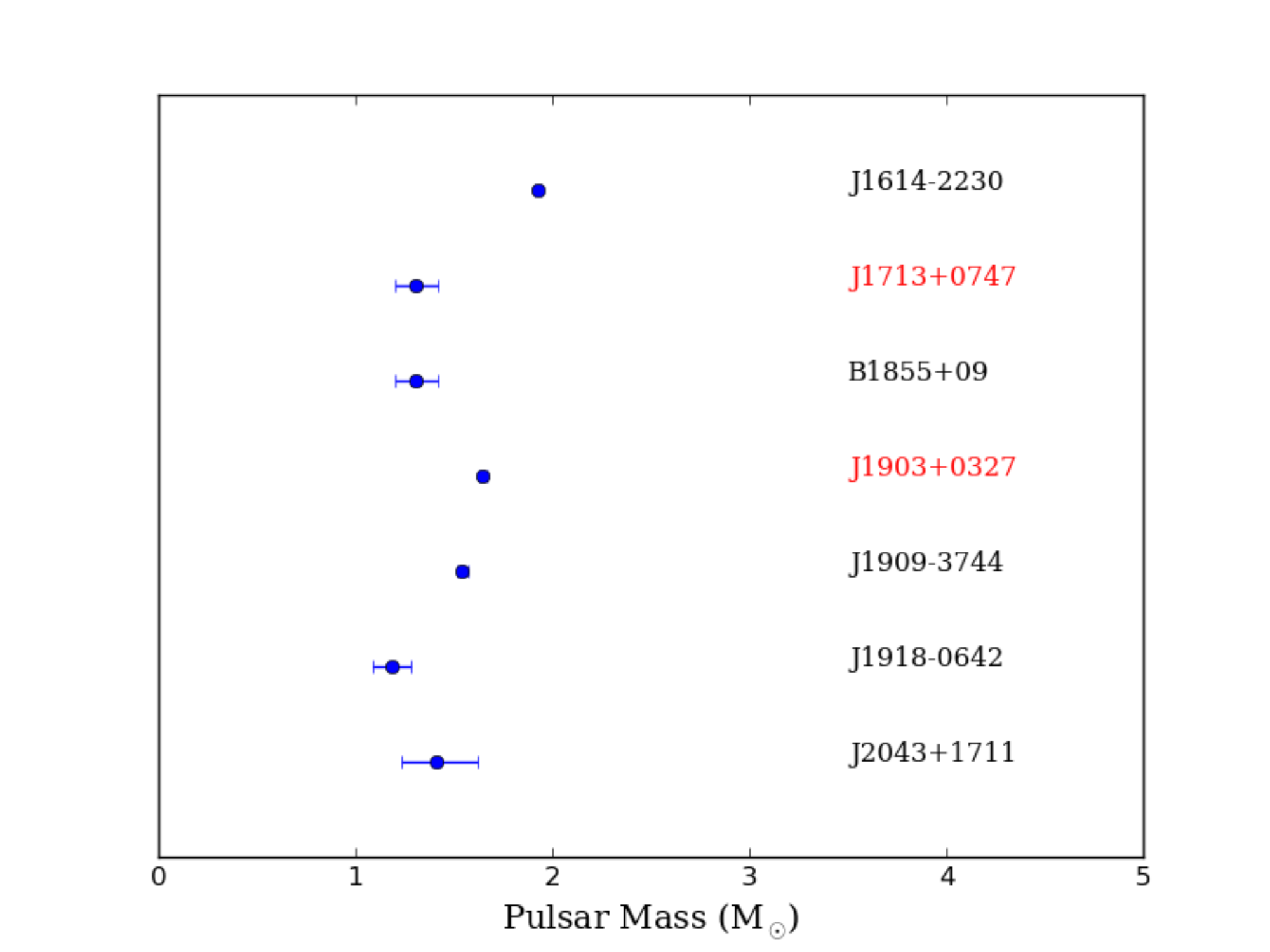} 
    \caption{Estimates of $m_{\rm p}$ for NANOGrav binary MSPs with the most significant Shapiro timing delays, which we define as  estimates of $m_{\rm c}$ with 68.3\% credible intervals that are 20\% of the median value. Red labels denote estimates obtained from $\chi^2$ grids that used the statistical significance of any observed secular variations as constraints, while black labels did not model variations in terms of mass or geometry. The blue points are median values and ranges are 68.3\% credible intervals derived from posterior PDFs obtained from using the traditional ($m_{\rm c}$, $\sin i$) parametrization of the Shapiro timing delay.}
    \label{fig:pulsarmasses}
\end{figure}

We have derived estimates of binary component masses and inclination angles for fourteen NANOGrav MSP-binary systems with significant measurements of the Shapiro timing delay. Four of these fifteen Shapiro-delay signals -- in PSRs J0613$-$0200,  J2017+0603, J2302+4442, and J2317+1439 -- have been measured for the first time. From the Shapiro timing delay alone, we were able to measure high-precision neutron star masses as low as $m_{\rm p} = 1.18^{+0.10}_{-0.09}\text{ M}_{\odot}$ for PSR J1918$-$0642 and as high as $m_{\rm p} = 1.928^{+0.017}_{-0.017}\text{ M}_{\odot}$ for PSR J1614$-$2230. Measurements of previously observed $\Delta_{\rm S}$ signals in the J1918$-$0642 and J2043+1711 systems have been significantly improved upon in this work, with the pulsar mass for PSR J2043+1711 $m_{\rm p} = 1.41^{+0.21}_{-0.18}\text{ M}_{\odot}$ being measured significantly for the first time. For the fourteen MSPs with significant $\Delta_{\rm S}$, we performed a rigorous analysis of the $\chi^2$ space for the two Shapiro-delay parameters, using priors uniform in the traditional ($m_{\rm c}$, $\sin i$) and orthometric ($h_3$, $\varsigma$) parametrizations of the Shapiro timing delay, in order to determine robust credible intervals of the physical parameters. We show the estimates of $m_{\rm p}$ for the most significant Shapiro timing delays in Figure \ref{fig:pulsarmasses}.

Most of the NANOGrav binary MSPs exhibit significant changes in one or more of their orbital elements over time. Whenever possible, we used the statistical significance of the observed orbital variations to further constrain the parameters of the observed Shapiro timing delay when performing the $\chi^2$-grid analysis. Assuming the validity of GR, we further constrained the component masses in the PSR J1600$-$3053 and PSR J1903+0327 systems, which both experience significant periastron advance due to strong-field gravitation; the precision of our $\dot{\omega}$ measurement for PSR J1903+0327 contributed to a highly constrained estimate of $m_{\rm p} = 1.65^{+0.02}_{-0.02}\text{ M}_{\odot}$ that is consistent with previous timing studies of this MSP using an independent data set. We also used the highly-significant $\dot{x}$ measurement in the PSR J1741+1351 system in combination with the Shapiro timing delay observed in this system, which allowed for an estimation of $\Omega$, albeit with a large uncertainty. We show the constrained  estimates of $m_{\rm p}$ in Figure \ref{fig:pulsarmasses} with red labels.

\begin{figure}
    \centering
    \includegraphics[scale=0.6]{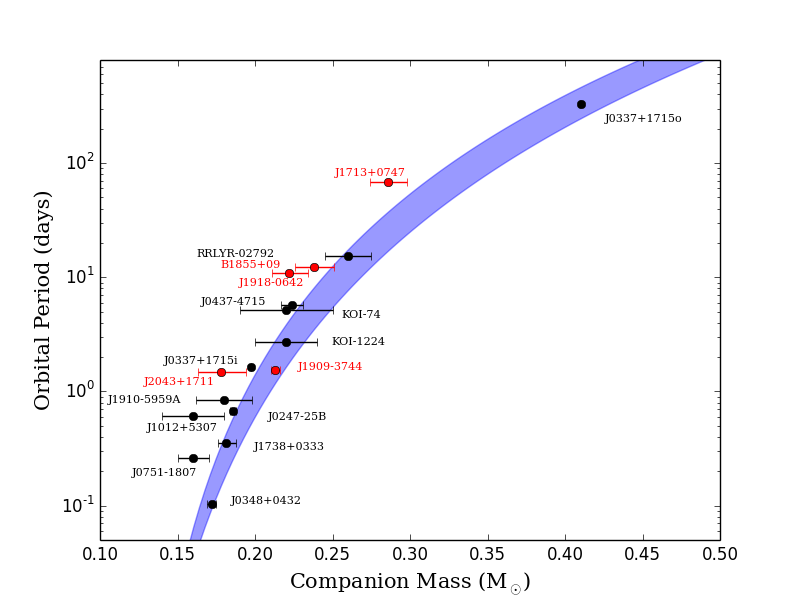}
    \caption{$P_{\rm b}$ versus $m_{\rm c}$ for binary systems with He-WD companions. Red points are our new measurements (see Figure 9). Black points are WD-mass measurements made for systems listed in Table \ref{tab:mcpb}. The shaded blue region is the expected correlation between $m_{\rm c}$ and $P_{\rm b}$, computed by \citet{ts99a}, for post-transfer He-WD binary systems with progenitor companions that have metallicities within the range $0.001 < Z < 0.02$.}
    \label{fig:mcPb}
\end{figure}

The relativistic Shapiro timing delay provides a direct measurement of the companion mass that is independent of the given system's evolutionary history, and that therefore can be used to test the plausibility of available binary-evolution paradigms. Figure \ref{fig:mcPb} illustrates the $P_{\rm b}$-vs-$m_{\rm c}$ estimates for the NANOGrav MSP-binary systems shown in Figure \ref{fig:pulsarmasses} that are known or suspected to have He-WD companions, as well as a blue-shaded region that corresponds to the theoretical $m_{\rm c}$-$P_{\rm b}$ correlation as predicted by \citet{ts99a}. PSR J1903+0327 is excluded since its companion is likely a main sequence star, while PSR J1614$-$2230 is excluded since its companion is a carbon-oxygen WD and is believed to have evolved through a different formation channel \citep{tlk11}. Figure \ref{fig:mcPb} is recreated from the one presented by \citet{tv14}. Black points denote precise measurements of $m_{\rm WD}$ in WD-binary systems examined in previous works; values and references for these data are provided in Table \ref{tab:mcpb}. The width of the shaded region represents possible correlated values of $P_{\rm b}$ and $m_{\rm WD}$ for progenitor donor stars with different chemical compositions, particularly with metallicities ($Z$) in the range $0.001 < Z < 0.02$. While our $m_{\rm c}$ estimates generally agree with the predicted correlation, additional measurements at higher companion masses are needed in order to perform a robust exploration of the correlation parameters and their credible intervals. 

The distribution of neutron-star masses can be directly inferred from available measurements of the Shapiro timing delay. Recent work has shown that an increasing number of these measurements can help delineate the roles of different supernovae processes in the formation of double-neutron-star binary systems \citep[e.g.][]{spr10} and assess the possible range of component masses for such systems \citep[e.g.][]{msf+15}, as well as derive the statistics for pulsar-binary populations that have evolved along different post-supernova evolutionary paths \citep{opns12, kkdt13}. In our study, the significant estimates of $m_{\rm p}$ span a range of $1.2-1.95\text{ M}_{\odot}$ in neutron star mass. PSRs J1614$-$2230 and J1918$-$0642 are at the high and low ends of our overall mass distribution, respectively. 

At its current level of precision, the low mass of PSR J1918$-$0642 is interesting since this MSP possesses spin parameters that are indicative of an old neutron star that experienced significant mass transfer and a substantial spin-up phase. The implication of a low ``birth mass" for neutron stars is consistent with early estimates of the initial-mass function \citep[e.g.][]{tww96}, though suggests that the neutron-star progenitor to J1918$-$0642 may have undergone an electron-capture supernova event \citep[e.g.][]{spr10} which produces comparatively less-massive neutron stars. Similar conclusions have been drawn for the lighter neutron stars in the J0737$-$3039A/B \citep{fsk+10} and J1756$-$2251 \citep{fsk+14} double-neutron-star binary systems, though the evolutionary history of these systems (with lesser degrees of mass transfer) are understood to be different than that expected for PSR J1918$-$0642.

Extending the data sets of these MSPs will refine observed secular variations due to PK and/or kinematic-bias effects within the next few years. Furthermore, extending TOA coverage in orbital phase for PSRs J0613$-$0200, J1949+3106, J2017+0603, J2302+4442, and J2317+1439 will improve the significance of the Shapiro timing delay that we report in this study. In particular, additional TOAs collected for PSRs J1640+2224 and J2317+1439 will help in the assessment of their complex orbital behavior as seen in the NANOGrav nine-year data set for these systems. The combination of NANOGrav high-precision TOAs with archival data published in previous studies will provide more accurate timing models and a complete picture of the physical processes that affect the NANOGrav MSP orbits. \\ [10pt]

{\it Author Contributions.} This study is the product of work performed for the doctoral dissertations of EF (supervised by IHS) and TTP (supervised by SMR and PBD), as well as the dedicated work of many people. EF performed most of the binary system computational analysis and astrophysical interpretations, generated all figures and tables, and drafted the text. TTP developed the source list, wrote proposals, lead observations, and analyzed the impact of observations targeted at detection of Shapiro delay. JAE developed and implemented the timing-noise model used for all NANOGrav pulsars and made substantial contributions to the analysis of binaries specifically for this work. PBD developed and implemented reduction pipelines to produce the TOAs analyzed here.  All authors performed observations for the NANOGrav project and developed timing models and made additional contributions to the data set as described in \cite{abb+15b}.

We thank P. C. C. Freire for useful discussion, as well as C. Bassa and C. Ng for comments on the manuscript. The NANOGrav project receives support from National Science Foundation (NSF) PIRE program award number 0968296 and NSF Physics Frontier Center award number 1430284. NANOGrav research at UBC is supported by an NSERC Discovery Grant and Discovery Accelerator Supplement and the Canadian Institute for Advanced Research. Part of this research was carried out at the Jet Propulsion Laboratory, California Institute of Technology, under a contract with the National Aeronautics and Space Administration. TTP was a student at the National Radio Astronomy Observatory while this project was undertaken. MTL was partially supported by NASA New York Space Grant award number NNX15AK07H. The National Radio Astronomy Observatory is a facility of the NSF operated under cooperative agreement by Associated Universities, Inc. The Arecibo Observatory is operated by SRI International under a cooperative agreement with the NSF (AST-1100968), and in alliance with Ana G. M\'{e}ndez-Universidad Metropolitana, and the Universities Space Research Association.

\bibliographystyle{apj_new}
\bibliography{nanograv_9yr_binary}

\clearpage
\begin{turnpage}
\input{Kep_table}
\end{turnpage}
\clearpage
\global\pdfpageattr\expandafter{\the\pdfpageattr/Rotate 90}
\clearpage

\global\pdfpageattr\expandafter{\the\pdfpageattr/Rotate 0}

\input{postKep_table}

\input{SDresults_table_orig}

\input{incllim_table}

\input{mcpb_table}

\clearpage

\clearpage
\clearpage
\global\pdfpageattr\expandafter{\the\pdfpageattr/Rotate 90}
\clearpage

\global\pdfpageattr\expandafter{\the\pdfpageattr/Rotate 0}
\clearpage

\end{document}

%% file: Kep_table.tex

\begin{deluxetable}{lllllllll}
    \tabletypesize{\scriptsize}
    \tablecaption{Keplerian Elements for the binary MSPs in the NANOGrav Nine-year Data Release}
    \tablewidth{0in}
    \tablehead{\colhead{PSR} & \colhead{$x$ (lt-s)} & \colhead{$P_b$ (days)} & \colhead{$e$} & \colhead{$\omega$ (deg)} & \colhead{$T_0$ (MJD)} & \colhead{$\eta$} & \colhead{$\kappa$} & \colhead{$T_{\rm asc}$ (MJD)}}
    \startdata
J0023+0923 & \phn\phn 0.03484105(11) & \phn\phn 0.13879914244(4) & 0.000025(5) & \phn 82.0(12.0) & 56179.082(5) & \phn 0.000024(5) & \phn 0.000003(5) & 56179.08248997(8) \\
J0613$-$0200 & \phn\phn 1.0914422(5) & \phn\phn 1.198512556680(13) & 0.00000443(17) & \phn 35.0(3.0) & 54890.089(10) & \phn 0.0000026(2) & \phn 0.00000362(8) & 54889.991808565(12) \\
J1012+5307 & \phn\phn 0.5818176(6) & \phn\phn 0.60467271380(6) & 0.0000013(17) & \phn 75.0(75.0) & 54901.95(13) & \phn 0.0000012(16) & \phn 0.0000003(16) & 54901.95231605(11) \\
J1455$-$3330 & \phn 32.3622120(3) & \phn 76.174567474(14) & 0.00016965(2) & 223.458(6) & 55531.1454(14) & \ldots & \ldots & \ldots \\
J1600$-$3053 & \phn\phn 8.8016526(10) & \phn 14.348468(3) & 0.000173741(11) & 181.854(16) & 55419.1115(6) & \ldots & \ldots & \ldots \\
J1614$-$2230 & \phn 11.29119744(7) & \phn\phn 8.68661942171(9) & 0.000001333(8) & 175.9(4) & 55662.053(10) & \phn 0.000000096(9) & $-$0.000001330(7) & 55658.145347857(6) \\
J1640+2224 & \phn 55.329717(4) & 175.460597(13) & 0.00079725(2) & \phn 50.7313(15) & 54784.4707(7) & \ldots & \ldots & \ldots \\
J1643$-$1224 & \phn 25.0725904(3) & 147.01739554(4) & 0.000505752(18) & 321.849(2) & 54870.5948(8) & \ldots & \ldots & \ldots \\
J1713+0747\tablenotemark{a} & \phn 32.34242188(14) & \phn 67.82513826930(19) &  0.0000749402(6) & 176.1963(16) &  53761.0327(3) & \ldots & \ldots & \ldots \\
J1738+0333 & \phn\phn 0.3434297(2) & \phn\phn 0.35479073425(6) & 0.0000004(10) & 252.0(140.0) & 55598.94(14) & $-$0.0000004(10) & $-$0.0000001(9) & 55598.93613993(12) \\
J1741+1351 & \phn 11.0033168(4) & \phn 16.3353478266(6) & 0.00000998(2) & 204.00(17) & 55812.321(8) & $-$0.00000406(3) & $-$0.00000912(2) & 55819.25468493(3) \\
J1853+1303 & \phn 40.76952255(13) & 115.653786432(6) & 0.000023700(6) & 346.656(11) & 56128.563(3) & \ldots & \ldots & \ldots \\
B1855+09 & \phn\phn 9.2307805(2) & \phn 12.32717119133(19) & 0.00002163(2) & 276.54(5) & 54975.5129(19) & \ldots & \ldots & \ldots \\
J1903+0327 & 105.593463(3) & \phn 95.17411738(8) & 0.43667843(2) & 141.6536021(15) & 55776.9743424(3) & \ldots & \ldots & \ldots \\
J1909$-$3744 & \phn\phn 1.89799095(4) & \phn\phn 1.533449451246(8) & 0.000000092(13) & 179.0(13.0) & 54514.49(6) & \phn 0.00000000(2) & $-$0.000000092(12) & 54513.989936084(3) \\
J1910+1256 & \phn 21.1291025(2) & \phn 58.466742058(5) & 0.00023020(2) & 106.013(6) & 54956.3186(11) & \ldots & \ldots & \ldots \\
J1918$-$0642 & \phn\phn 8.3504665(2) &  \phn 10.9131775801(2) & 0.000020340(18) & 219.38(6) & 54893.7305(17) & $-$0.00001291(2) & $-$0.000015721(13) & 54897.63652454(2) \\
J1949+3106 & \phn\phn 7.288647(7) & \phn\phn 1.9495344177(8) & 0.0000429(3) & 208.0(6) & 56365.552(3) & $-$0.0000201(5) & $-$0.0000379(2) & 56365.97423581(3) \\
B1953+29 & \phn 31.4126915(2) & 117.349097292(19) & 0.000330230(15) & \phn 29.483(2) & 55265.7096(7) & \ldots & \ldots & \ldots \\
J2017+0603 & \phn\phn 2.1929208(9) & \phn\phn 2.1984811364(4) & 0.00000685(15) & 177.0(3.0) & 56201.626(15) & \phn 0.0000004(3) & $-$0.00000684(15) & 56200.64259488(3) \\
J2043+1711 & \phn\phn 1.6239584(2) & \phn\phn 1.48229078649(14) & 0.00000489(13) & 240.4(1.2) & 56173.974(5) & $-$0.00000425(13) & $-$0.00000242(9) & 56174.306240718(10) \\
J2145$-$0750 & \phn 10.16410849(17) & \phn\phn 6.83890250963(11) & 0.000019295(19) & 200.91(5) & 54902.6174(9) & \ldots & \ldots & \ldots \\
J2214+3000 & \phn\phn 0.0590817(3) & \phn\phn 0.4166329463(9) & 0.000009(11) & 345.0(72.0) & 56221.96(8) & $-$0.000002(10) & \phn 0.000008(11) & 56221.9632381(4) \\
J2302+4442 & \phn 51.4299676(5) & 125.93529697(13) & 0.000503021(17) & 207.8925(18) & 56302.6599(6) & \ldots & \ldots & \ldots \\
J2317+1439 & \phn\phn 2.313943(4) & \phn\phn 2.45933146519(2) & 0.0000007(5) & 101.0(42.0) & 54976.1(3) & $-$0.0000007(5) & \phn 0.00000015(6) & 54976.609358785(14) \\
    \enddata
    \label{tab:binarypar}
    \tablecomments{Values in parentheses denote the $1\sigma$ uncertainty in the preceding digit(s), as determined from TEMPO2. For MSPs with both DD and ELL1 parameters listed in this table, we used the ELL1 model to describe the Keplerian orbit in the TEMPO2 fit, and then derived the corresponding DD values; the 1$\sigma$ uncertainties for the derived DD parameters were computed by propagating $1\sigma$ uncertainties in the fitted ELL1 parameters.}
    \tablenotetext{a}{The values for PSR J1713+0747 were taken from \citet{zsd+15}.}
\end{deluxetable}

%% file: postKep_table.tex
\begin{deluxetable}{lllllllcc}
    \tabletypesize{\scriptsize}
    \tablecaption{Secular Variations and Shapiro-Delay Parameters in the NANOGrav nine-year Data Release}
    \tablewidth{0in}
    \tablehead{\colhead{PSR} & \colhead{$\dot{\omega}$ (deg yr$^{-1}$)} & \colhead{$\dot{x}$ (10$^{-12}$)} & \colhead{$\dot{P}_{\rm b}$ (10$^{-12}$)} & \colhead{$h_3$ ($\mu$s)} & \colhead{$h_4$ ($\mu$s)} & \colhead{$\varsigma$} & \colhead{Detection of $\Delta_{\rm S}$?} & \colhead{Span (yr)}}
    \startdata
J0023+0923 & \ldots & \ldots & \ldots & \phn 0.06(5) & $-$0.00(6) & \ldots & N & 2.3 \\
J0613$-$0200 & \ldots & \ldots & \ldots  & \phn 0.28(3) & \ldots & 0.74(8) & Y & 8.6 \\
J1012+5307 & \ldots & \ldots & \ldots & $-$0.00(9) & \phn 0.05(10) & \ldots & N & 9.2 \\
J1455$-$3330 & \ldots & $-$0.021(5) & \ldots & \phn 0.3(2) & \ldots & 0.7(4) & N & 9.2 \\
J1600$-$3053 & \phn 0.007(2) & $-$0.0017(9)\tablenotemark{a} & \ldots & \phn 0.39(3) & \ldots & 0.62(6) & Y & 6.0 \\
J1614$-$2230 & \ldots & \ldots & 1.3(7)\tablenotemark{a} & \phn 2.329(11) & \ldots & 0.9859(2) & Y & 5.1 \\
J1640+2224 & $-$0.00028(5) & \phn 0.0145(10) & \ldots & \phn 0.57(6) & \ldots & 0.61(8) & Y & 8.9 \\
J1643$-$1224 & \ldots & $-$0.047(3) & \ldots & $-$0.09(13) & \ldots & 1.2(8) & N & 9.0 \\
J1713+0747 & \ldots & \phn 0.00645(11) & \ldots & \phn 0.54(3) & \ldots & 0.73(1) & Y\tablenotemark{b} & 8.8 \\
J1738+0333 & \ldots & \ldots & \ldots & \phn 0.02(12) & \phn 0.06(13) & \ldots & N & 4.0 \\
J1741+1351 & \ldots & $-$0.0094(18) & \ldots & \phn 0.46(6) & \ldots & 0.85(10) & Y & 4.2 \\
J1853+1303 & \ldots & \phn 0.0147(19) & \ldots & \phn 0.11(11) & \ldots & 0.5(1.2) & N & 5.6 \\
B1855+09 & \ldots & \ldots & \ldots & \phn 1.04(4) & \ldots & 0.969(5) & Y & 8.9 \\
J1903+0327 & \phn 0.0002410(13) & \ldots & \ldots & \phn 2.0(3) & \ldots & 0.70(8) & Y & 4.0 \\
J1909$-$3744 & \ldots & $-$0.00044(16)\tablenotemark{a} & 0.509(9) & \phn 0.868(7) & \ldots & 0.9381(16) & Y & 9.1 \\
J1910+1256 & \ldots & $-$0.017(2) & \ldots & \phn 0.3(2) & \ldots & 0.7(7) & N & 8.8 \\
J1918$-$0642 & \ldots & \ldots & \ldots & \phn 0.83(3) & \ldots & 0.918(8) & Y & 9.0 \\
J1949+3106 & \ldots & \ldots & \ldots & \phn 2.5(5) & \ldots & 0.77(10) & Y & 1.2 \\
B1953+29 & \ldots & \phn 0.011(3) & \ldots & $-$0.1(6) & \ldots & 0.8(5) & N & 7.2 \\
J2017+0603 & \ldots & \ldots & \ldots & \phn 0.31(7) & \ldots & 0.72(8) & Y & 1.7 \\
J2043+1711 & \ldots & \ldots & \ldots & \phn 0.60(3) & \ldots & 0.890(13) & Y & 2.3 \\
J2145$-$0750 & \ldots & \phn 0.0098(19) & \ldots & \phn 0.10(5) & \ldots & 0.94(17) & N & 9.1 \\
J2214+3000 & \ldots & \ldots & \ldots & $-$0.3(2) & $-$0.1(3) & \ldots & N & 2.1 \\
J2302+4442 & \ldots & \ldots & \ldots & \phn 1.5(3) & \ldots & 0.55(15) & Y & 1.7 \\
J2317+1439\tablenotemark{c} & \ldots & \ldots & \ldots & \phn 0.33(6) & \ldots & 0.49(14) & Y & 8.9
    \enddata
    \label{tab:postKeppar}
    \tablecomments{Values in parentheses denote the $1\sigma$ uncertainty in the preceding digit(s), as determined from TEMPO2.}
    \tablenotetext{a}{These parameters did not pass the F-test criterion for inclusion into the NANOGrav data-release timing solutions but were nonetheless fitted for; see the second paragraph of Section \ref{sec:analysis} for details.}
    \tablenotetext{b}{The values for PSR J1713+0747 were taken from \citet{zsd+15}, and the orthometric parameters were computed from their reported traditional ($m_{\rm c}$, $\sin i$) estimates. We do not analyze this MSP any further, and only focus on the other 14 NANOGrav binary MSPs in this work.}
    \tablenotetext{c}{We altered the timing model for PSR J2317+1439 in order to remove the ($\dot{\eta}$, $\dot{\kappa}$) parameters used in the NANOGrav nine-year timing model for this pulsar, and instead model the Shapiro timing delay; see Section \ref{subsec:J2317} for a discussion.}
\end{deluxetable}

%% file: SDresults_table_orig.tex
\begin{deluxetable}{l|ll|ll|ll}
    \tabletypesize{\scriptsize}
    \tablecaption{Estimates of Shapiro-delay Parameters from $\chi^2$-grid Analyses}
    \tablewidth{0in}
    \tablehead{\multicolumn{1}{c}{PSR} & \multicolumn{2}{|c|}{\underline{Pulsar Mass (M$_{\odot})$}} & \multicolumn{2}{|c|}{\underline{Companion Mass (M$_{\odot}$)}} & \multicolumn{2}{|c|}{\underline{System Inclination (deg)}} \\ [5pt] & Trad & Ortho & Trad & Ortho & Trad & Ortho}
    \startdata
    J0613$-$0200 & $2.3^{+2.7}_{-1.1}$ & $2.1^{+2.1}_{-1.0}$ & $0.21^{+0.23}_{-0.10}$ & $0.19^{+0.15}_{-0.07}$ & $66^{+8}_{-12}$ & $68^{+7}_{-10}$ \\ [5pt]
    J1600$-$3053\tablenotemark{a} & $2.4^{+1.5}_{-0.9}$ & $2.4^{+1.3}_{-0.8}$ & $0.33^{+0.14}_{-0.10}$ & $0.33^{+0.13}_{-0.08}$ & $63^{+5}_{-5}$ & $64^{+4}_{-5}$ \\ [5pt]
    J1614$-$2230 & $1.928^{+0.017}_{-0.017}$ & $1.928^{+0.017}_{-0.017}$ & $0.493^{+0.003}_{-0.003}$ & $0.493^{+0.003}_{-0.003}$ & $89.189^{+0.014}_{-0.014}$ & $89.188^{+0.014}_{-0.014}$ \\ [5pt]
    J1640$+$2224 & $4.4^{+2.9}_{-2.0}$ & $5.2^{+2.6}_{-2.0}$ & $0.6^{+0.4}_{-0.2}$ & $0.7^{+0.3}_{-0.2}$ & $60^{+6}_{-6}$ & $58^{+6}_{-6}$ \\ [5pt]
    J1713$+$0747\tablenotemark{a,b} & $1.31^{+0.11}_{-0.11}$ & $1.31^{+0.11}_{-0.11}$ & $0.286^{+0.012}_{-0.012}$ & $0.286^{+0.012}_{-0.012}$ & $71.9^{+0.7}_{-0.7}$ & $71.9^{+0.7}_{-0.7}$  \\ [5pt]
    J1741$+$1351\tablenotemark{a} & $1.87^{+1.26}_{-0.69}$ & $1.78^{1.08}_{-0.63}$ & $0.32^{+0.15}_{-0.09}$ & $0.31^{+0.13}_{-0.08}$ & $66^{+5}_{-6}$ & $66^{+5}_{-6}$ \\ [5pt]
    B1855$+$09 & $1.30^{+0.11}_{-0.10}$ & $1.31^{+0.12}_{-0.10}$ & $0.236^{+0.013}_{-0.011}$ & $0.238^{+0.013}_{-0.012}$ & $88.0^{+0.3}_{-0.4}$ & $88.0^{+0.3}_{-0.4}$ \\ [5pt]
    J1903$+$0327\tablenotemark{a} & $1.65^{+0.02}_{-0.02}$ & $1.65^{+0.02}_{-0.03}$ & $1.06^{+0.02}_{-0.02}$ & $1.06^{+0.02}_{-0.02}$ & $72^{+2}_{-3}$ & $72^{+2}_{-3}$ \\ [5pt]
    J1909$-$3744 & $1.55^{+0.03}_{-0.03}$ & $1.55^{+0.03}_{-0.03}$ & $0.214^{+0.003}_{-0.003}$ & $0.214^{+0.003}_{-0.003}$ & $86.33^{+0.09}_{-0.10}$ & $86.33^{+0.09}_{-0.10}$ \\ [5pt]
    J1918$-$0642 & $1.18^{+0.10}_{-0.09}$ & $1.19^{+0.10}_{-0.09}$ & $0.219^{+0.012}_{-0.011}$ & $0.219^{+0.012}_{-0.011}$ & $85.0^{+0.5}_{-0.5}$ & $85.0^{+0.5}_{-0.5}$ \\ [5pt]
    J1949$+$3106 & $4.0^{+3.6}_{-2.5}$ & $4.0^{+3.4}_{-2.3}$ & $2.1^{+1.6}_{-1.0}$ & $1.9^{+1.5}_{-0.9}$ & $67^{+9}_{-8}$ & $68^{+8}_{-8}$ \\ [5pt]
    J2017$+$0603 & $2.4^{+3.4}_{-1.4}$ & $2.0^{+2.8}_{-1.1}$ & $0.32^{+0.44}_{-0.16}$ & $0.27^{+0.30}_{-0.12}$ & $62^{+9}_{-12}$ & $65^{+7}_{-11}$ \\ [5pt]
    J2043$+$1711 & $1.41^{+0.20}_{-0.18}$ & $1.43^{+0.21}_{-0.18}$ & $0.175^{+0.016}_{-0.015}$ & $0.177^{+0.017}_{-0.015}$ & $83.2^{+0.8}_{-0.9}$ & $83.1^{+0.8}_{-0.9}$ \\ [5pt]
    J2302$+$4442 & $5.3^{+3.2}_{-3.6}$ & $5.5^{+3.0}_{-3.2}$ & $2.3^{+1.7}_{-1.3}$ & $1.8^{+1.6}_{-1.0}$ & $54^{+12}_{-7}$ & $57^{+11}_{-9}$ \\ [5pt]
    J2317$+$1439 & $4.7^{+3.4}_{-2.8}$ & $4.1^{+3.5}_{-2.4}$ & $0.7^{+0.5}_{-0.4}$ &$0.5^{+0.5}_{-0.3}$  & $47^{+10}_{-7}$ & $51^{+10}_{10}$ 
    \enddata
    \label{tab:SDresults}
    \tablecomments{All uncertainties reflect 68.3\% credible intervals. ``Trad" refers to estimates made with the traditional ($m_{\rm c}$, $\sin i$) Shapiro-delay model, while ``Ortho" refers to those made with the orthometric ($h_3$, $\varsigma$) model. Difference in median values and credible intervals reflect the consequence in choosing uniform prior PDFs on the ($m_{\rm c}$, $\sin i$) or ($h_3$, $\varsigma$) parameters for weak measurements of $\Delta_{\rm S}$.}
    \tablenotetext{a}{The observed secular variations in this system were used to constrain the Shapiro-delay parameters.}
    \tablenotetext{b}{The Shapiro-delay estimates for PSR J1713+0747 were taken from \citet{zsd+15}, which used the NANOGrav nine-year data set as well as historical TOAs collected for previous studies.}
\end{deluxetable}

%% file: incllim_table.tex
\begin{deluxetable}{lcc}

\tabletypesize{\scriptsize}
\tablecaption{Limits on Inclination from Non-detection of Shapiro Delay and/or Detection of $\dot{x}$}
\tablewidth{0in}
\tablehead{\colhead{PSR} & \colhead{$i_{\rm SD}$ (deg)} & \colhead{$i_{\dot{x}}$ (deg)}}
J0023$+$0923 & < 56 & \ldots \\
J1012$+$5307 & < 66 & \ldots \\
J1455$-$3330 & < 85 & < 77 \\
J1643$-$1224 & < 73 & < 37 \\
J1738$+$0333 & < 70 & \ldots \\
J1853$+$1303 & < 74 & < 63 \\
J1910$+$1256 & < 63 & < 63 \\
B1953$+$29 & < 80 & < 77 \\
J2145$-$0750 & < 80 & < 73 \\
J2214$+$3000 & < 75 & \ldots 
\tablecomments{All upper limits are at 95\% confidence.}
\label{tab:upperlim}
\end{deluxetable}

%% file: mcpb_table.tex
\begin{deluxetable}{llcc}
    \tabletypesize{\scriptsize}
    \tablecaption{Precise Estimates of $m_{\rm c}$ for Low-Mass He White Dwarfs}
    \tablewidth{0in}
    \tablehead{\colhead{System Name} & \colhead{$m_{\rm c}$ (M$_{\odot}$)} & \colhead{$P_{\rm b}$ (days)} & \colhead{References (Mass Measurement, Identification)}}
    \startdata
    PSR J0348+0432 & 0.172(3) & \phn\phn 0.102 & \citet{afw+13} \\
    PSR J0751$-$1807 & 0.16(1) & \phn\phn 0.263 & \citet{bvk06}, \citet{dcl+16} \\
    PSR J1738+0333 & $0.181^{+0.007}_{-0.005}$ & \phn\phn 0.354 & \citet{avk+12} \\
    PSR J1012+5307 & 0.16(2) & \phn\phn 0.604 & \citet{vbk96}, \citet{vbjj05} \\ 
    J0247$-$25B & 0.186(2) & \phn\phn 0.667 & \citet{msm+13} \\ 
    PSR J1910$-$5959A & 0.180(18) & \phn\phn 0.837 & \citet{bvkv06}, \citet{cbp+12} \\ 
    PSR J0337+1715i & 0.19751(15) & \phn\phn 1.629 & \citet{rsa+14}, \citet{kvk+14} \\ 
    KOI 1224 & 0.22(2) & \phn\phn 2.698 & \citet{brvc12} \\
    KOI 74 & 0.22(3) & \phn\phn 5.189 & \citet{vrb+10} \\ 
    PSR J0437$-$4715 & 0.224(7) & \phn\phn 5.741 & \citet{dkp+12}, \citet{rhc+16} \\ 
    RRLYR 02792 & 0.260(15) & \phn 15.243 & \citet{ptg+12} \\ 
    PSR J0337+1715o & 0.4101(3) & 327.257 & \citet{rsa+14}
    \enddata
    \label{tab:mcpb}
    \tablecomments{Uncertainties in $P_{\rm b}$ are suppressed due to the high precision to which they are measured. Values in parentheses denote the $1\sigma$ uncertainty in the preceding digit(s).}
\end{deluxetable}